\documentclass[11pt,a4paper]{article}
\pdfoutput=1
\usepackage{jheppub}
\usepackage{graphicx}
\usepackage{amsmath}
\usepackage{verbatim}
\usepackage{amssymb}
\usepackage{inputenc, array}
\usepackage{textcomp}
\usepackage{appendix}
\usepackage{mathrsfs}
\usepackage[dvipsnames]{xcolor}
\newcommand{\e}{\epsilon}
\newcommand{\be}[1]{\begin{equation}\label{#1} }
\newcommand{\ee}{\end{equation}}
\newcommand{\bea}[1]{\begin{eqnarray}\label{#1} }
\newcommand{\eea}{\end{eqnarray}}
\newcommand{\p}{\partial}
\newcommand{\refb}[1]{(\ref{#1})}

\renewcommand{\L}{{\mathcal{L}}}

\newcommand{\bL}{\bar{{\mathcal{L}}}}

\renewcommand{\(}{\left(}
\renewcommand{\)}{\right)}

\renewcommand{\a}{\alpha}

\renewcommand{\b}{\beta}
\renewcommand{\t}{\tau}
\newcommand{\s}{\sigma}

\newcommand{\ie}{\emph{i.e.}}
\newcommand{\eg}{\emph{e.g.}}

\title{Carroll covariant scalar fields in two dimensions}
\author[a]{Arjun Bagchi,} \author[b]{Aritra Banerjee,}  \author[a]{Sudipta Dutta,} \author[a]{Kedar S.~Kolekar,} \author[a]{Punit Sharma.}
\author{\\}

\affiliation[a]{Indian Institute of Technology Kanpur, Kanpur 208016, INDIA.\\} 

\affiliation[b]{Okinawa Institute of Science \& Technology, 1919-1 Tancha, Onna-son, Okinawa 904-0495, JAPAN.\\}

\emailAdd{(abagchi, dsudipta, kedarsk, spunit)@iitk.ac.in, aritra.banerjee@oist.jp}

\preprint{}

\abstract{Conformal Carroll symmetry generically arises on null manifolds and is important for holography of asymptotically flat spacetimes, generic black hole horizons and tensionless strings. In this paper, we focus on two dimensional ($2d$) null manifolds and hence on the $2d$ Conformal Carroll or equivalently the $3d$ Bondi-Metzner-Sachs (BMS) algebra. Using Carroll covariance, we write the most general free massless Carroll scalar field theory and discover {\em{three inequivalent actions}}. Of these, two viz. the time-like and space-like actions, have made their appearance in literature before. We uncover a third that we call the mixed-derivative theory. As expected, all three theories enjoy off-shell BMS invariance. Interestingly, we find that the on-shell symmetry of mixed derivative theory is a single Virasoro algebra instead of the full BMS. We discuss potential applications to tensionless strings and flat holography.}

\begin{document}
\maketitle

%%%%%%%%%%%%%%%%%%%%%%%%%%%%%%%%%%%%%%%%
%%%%%%%%%%%% INTRODUCTION %%%%%%%%%%%%%%

\section{Introduction}

Null surfaces are essential for understanding many aspects of gravitational physics and are crucial for quantum gravity as well. The most important of these null manifolds are the boundary of asymptotically flat spacetime $\mathscr{I}^\pm$ and the event horizon of generic black holes. In asymptotically flat spacetimes, the asymptotic in and out states are defined on $\mathscr{I}^\pm$ and hence all scattering processes are linked to this inexorably. $S$-matrix elements are the only observables for a theory of quantum gravity in flatspace and hence the understanding of the null boundary is vital. 

\medskip

The event horizons of black holes on the other hand, provide a description of the degrees of freedom of the black holes themselves. The Bekenstein-Hawking formula, famously, gives the entropy of a black hole, not in terms of its volume, but the area of the event horizon. The seeds of the Holographic Principle are in this above statement. So for a better understanding of black holes, and indeed the Holographic Principle in general, understanding the nature of the event horizon is essential. 

\medskip

Null surfaces are endowed with degenerate metrics. In modern parlance, null surfaces are described by what is called a \textit{Carrollian} structure.  A $(d+1)$ dimensional Carroll manifold is a fibre-bundle with spatial part forming a $d$-dimensional base space and the null direction forming the $1d$ fibre. 

\medskip

\begin{figure}[h!]\label{lightcone}
\begin {center}
    \includegraphics [scale = 0.25] {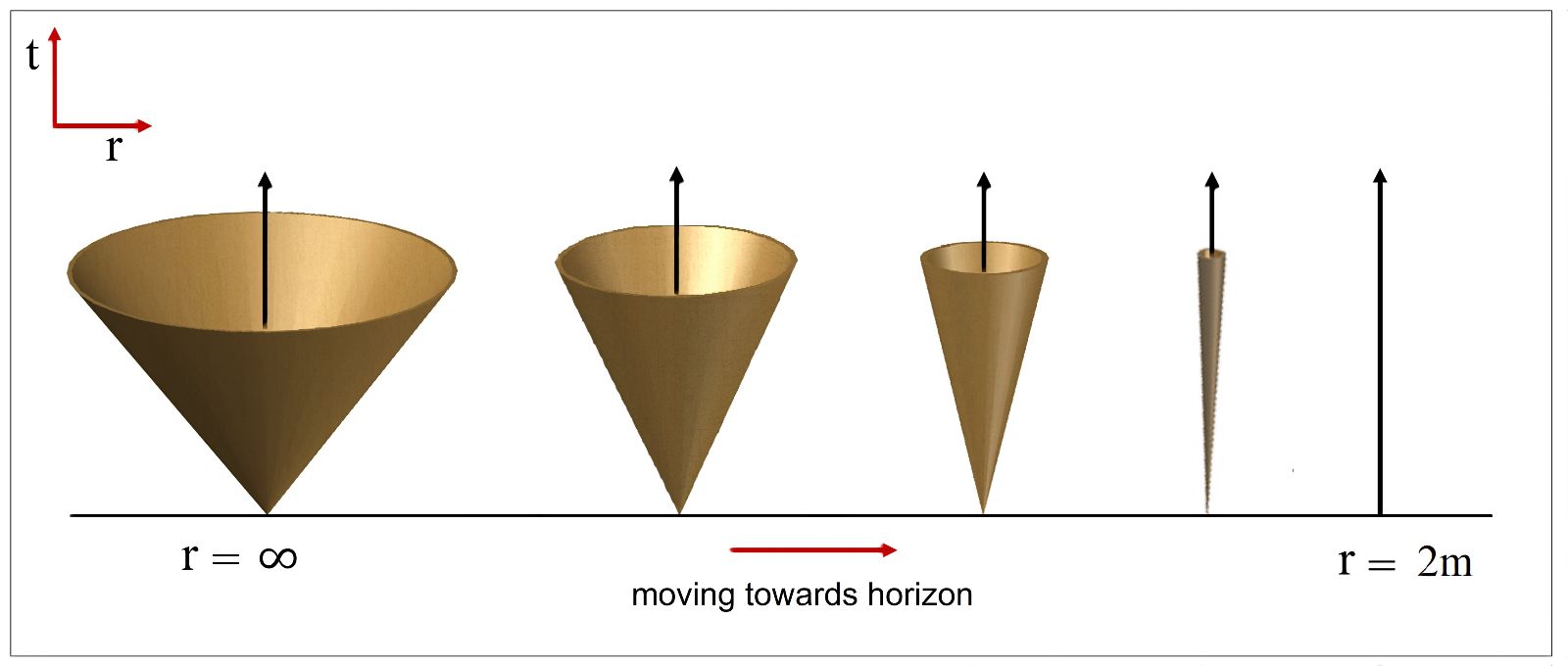} \ \
\caption{Collapsing of the lightcone towards the Schwarzschild horizon.}
\end {center}
\end{figure}

\medskip
Carroll manifolds can be obtained by taking a Carroll or an ultrarelativistic (UR) limit $c\rightarrow 0$, \ie\ sending the speed of light to zero, on (pseudo-) Riemannian manifolds \cite{LBLL,Henneaux:1979vn,Duval:2014uoa,Duval:2014uva,Duval:2014lpa}. A particularly useful way of visualising this is to think of the familiar Schwarzschild black hole in the usual Schwarzschild coordinates:
\be{}
ds^2 = - \left(1-\frac{2m}{r}\right) dt^2 + \frac{dr^2}{\left(1-\frac{2m}{r}\right)} + r^2 d\Omega^2. \nonumber
\ee
Here $m$ is the mass of the Schwartzschild black hole and $r=2m$ is the event horizon. Let us focus on lightcones and neglect the angular direction at the moment. The slope of the lightcone is given by 
\be{}
\frac{dr}{dt} = \pm \left(1- \frac{2m}{r}\right). \nonumber
\ee 
$r\to\infty$ of course gives one the usual flat spacetimes. The speed of light, in the natural units, in this case is $c=1$. The lightcone, as one expects, is at 45 degree angles. As one moves into the bulk, with decreasing $r$, the lightcone shrinks more and more. At $r=2m$, the horizon is hit. The lightcone closes up completely, the speed of light $c\to0$ and we get a Carrollian manifold endowed with a degenerate metric. See figure \ref{lightcone} for a depiction of this process. This is of course a coordinate dependent picture in the Schwarzschild background, but any generic Carrollian limit can be visualised in this way and the coordinate independent statement is that whenever a null surface is hit, the lightcones will close up and the effective velocity of light in these codimension-one surfaces will go to zero. 

\medskip

\subsection{Carroll and Asymptotically Flat Spacetime}

The asymptotic symmetries of flat spacetimes have been the cynosure of a lot of recent attention. It has been known since the 1960's following the seminal work of Bondi, van der Burgh, Metzner and Sachs \cite{Bondi:1,Sachs:1962zza}, that in $4d$  flat spacetimes, the asymptotic symmetry group on $\mathscr{I}^\pm$ extends beyond the expected Poincare group to the infinite dimensional BMS group. The relevant analysis in $3d$ also exhibits a similar behaviour. The BMS$_3$ algebra reads \cite{Barnich:2006av}
\begin{subequations}\label{bms3}
\bea{}
&& [L_n, L_m] = (n-m) L_{n+m} + \frac{c_L}{12} (n^3-n) \delta_{n+m,0} \\
&& [L_n, M_m] = (n-m) M_{n+m} + \frac{c_M}{12} (n^3-n) \delta_{n+m,0} \\
&& [M_n, M_m] = 0.
\eea
\end{subequations}
Here the generators $M_n$ generate angle-dependent translations of the null direction and are called supertranslations and $L_n$'s generate the $\text{Diff}(S^1)$ of the circle at infinity. We have already stressed that e.g. $\mathscr{I}^+$ is a null manifold and hence has intrinsic Carroll structures defined on it. So the question is how Carroll symmetries are related to BMS symmetries. We are interested in understanding theories that live on the null surface. So any degrees of freedom (d.o.f) should not leave the surface and this means that these d.o.f. necessarily have to be massless. We are hence led to considering conformal Carrollian theories or CFTs that live on Carroll manifolds. Conformal Carroll symmetries, following this intuition, turn out to be isomorphic to BMS symmetries and this connection can be proved rigourously. In a nutshell \cite{Duval:2014uva,Duval:2014lpa}
\be{}
\mathfrak{Ccarr}(d)\simeq \mathfrak{bms}(d+1),
\ee
or in words, the conformal Carroll algebra turns out to be isomorphic to the BMS algebra in one higher dimension 
{\footnote{The full story is a bit more intricate. The conformal isometries of a $d$ dimensional flat Carroll manifold, generically, generate the conformal Carroll algebra of level $N$, $\mathfrak{ccarr}_N(d)$.
The so called `dynamical exponent' $z=\frac{2}{N}$ captures the relative scaling between space and time.
For $N=2$, space and time dilate homogeneously and it is in this case that the conformal Carroll algebra turns out to be isomorphic to the BMS algebra in one higher dimension, i.e. $\mathfrak{ccarr}_2(d)\simeq \mathfrak{bms}(d+1)$ \cite{Duval:2014uva,Duval:2014lpa}.}}. 

\medskip

A word or two about the limiting procedure now. From an algebraic perspective, the Carroll contraction $c\rightarrow 0$ of the Poincar\'e group gives the Carroll group, which is the isometry group of the flat Carroll spacetime. We have said that the conformal Carroll group and its corresponding algebra are infinite dimensional. However, its relativistic counterpart for $d>2$ has only finite number of generators. Systematic contractions from relativistic conformal algebra yield only the global part of conformal Carroll algebra in dimensions $d\geq3$. There is a way of generating the infinite algebra by rewriting the finite algebra in a suggesting basis and lifting the generators to their infinite dimensional versions \cite{Bagchi:2012cy}.

\medskip

In this paper, we will be interested in $2d$ Carrollian and Conformal Carrollian structures and hence from the point of view of the limit, the relativistic system from which the Carroll Conformal algebra would be derived is of course infinite dimensional as well. The parent symmetry of interest is two copies of the Virasoro algebra. The algebra contracts into the BMS$_3$ algebra as \cite{Barnich:2006av, Bagchi:2012cy}
\be{}
\L_n - \bL_{-n} = L_n, \quad \L_n + \bL_{-n} = \frac{1}{\e} M_n. 
\ee
We will call this the ultra-relativistic (UR) contraction of the algebra. There is another contraction we will be interested in for the purposes of this paper. This is the non-relativistic (NR) contraction \cite{Bagchi:2009pe} which is 
\be{}
\L_n + \bL_{n} = L_n, \quad \L_n - \bL_{n} = \frac{1}{\e} M_n. 
\ee
This also yields the BMS$_3$ from two copies of Virasoro. This magic of two dimensions makes the NR and UR contractions give isomorphic algebras in the limit. This is because the corresponding Galilean and Carrollian algebras as well as their infinite dimensional conformal cousins are isomorphic in $d=2$ \cite{Bagchi:2010zz}. 

\medskip

We will elaborate especially on the geometric perspective of Carroll and conformal Carroll symmetries in two dimensions in this work. Below we discuss where conformal Carroll symmetries find uses in modern theoretical physics. 

\subsection{Flatspace Holography}
Borrowing popular wisdom from the AdS/CFT correspondence \cite{Maldacena:1997re}, we can postulate that matching of asymptotic symmetries of the bulk gravitational theory with the global symmetries of the dual boundary quantum field theory is the building block of the formulation of a holographic correspondence in a general spacetime. 

\medskip

For asymptotically flat spacetimes, the null boundaries are Carrollian manifolds, and the associated asymptotic symmetry group (ASG) is the BMS group \cite{Bondi:1,Sachs:1962zza,Barnich:2006av}. Following this line of logic, the formulation of holographic correspondence for asymptotically flat spacetimes should involve field theories living on null boundaries governed by the BMS group as global symmetries \cite{Barnich:2010eb,Bagchi:2010zz,Bagchi:2012cy}. 

\medskip

This approach, nowadays known as Carrollian holography, has indeed met with some success for holography for lower dimensional (specifically $3d$) asymptotically flat spacetimes. In $3d$ bulk with BMS$_3$ as the ASG, some concrete developments involve a proposal for a field theory dual \cite{Bagchi:2012yk}, matching of entropy \cite{Bagchi:2012xr,Barnich:2012xq,Bagchi:2013qva}, the computation of stress tensor correlators \cite{Bagchi:2015wna}, entanglement entropy \cite{Bagchi:2014iea,Jiang:2017ecm,Hijano:2017eii}, and more. A selected, non-exhaustive list of papers along these directions are \cite{Bagchi:2013lma, Detournay:2014fva, Afshar:2013vka, Gonzalez:2013oaa,Hartong:2015usd, Hartong:2015xda, Bagchi:2016geg, Barnich:2014cwa, Fareghbal:2014qga, Grumiller:2019xna, Ciambelli:2018wre} and for related higher dimensional stories, the reader can have a look at \cite{Bagchi:2016bcd}.

\medskip

Along another distinct avenue, particularly in four dimensions, following the initiative by Strominger and collaborators, a lot of progress have been made in linking up asymptotic symmetries of a $4d$ flat space with a dual $2d$ CFT living on the Celestial Sphere.  This approach, often called Celestial Holography, has generated a lot of interest and a lot of novel results for 4d scattering amplitudes and related asymptotic symmetries. For a glimpse of  recent developments one could look at the excellent reviews \cite{Strominger:2017zoo, Pasterski:2021rjz, Raclariu:2021zjz}. 

\medskip

More recently attempts have been made to construct a bridge between these two formulations \cite{Donnay:2022aba, Bagchi:2022emh} and it has been crucially shown inspired by ideas from Celestial holography that Carroll CFTs provide a concrete framework for understanding $S$-matrix elements \cite{Bagchi:2022emh}, thus developing a connection that was previously lacking. 

\medskip

Hence a road to holography in asymptotically flat spacetimes is crucially dependent on understanding conformal Carroll symmetry and the field theories which realise these symmetries. In this paper, we will be specifically interested in two dimensional theories and hence the algebra \refb{bms3}.

\subsection{Horizons, Carroll strings and other things}

Having discussed flat holography which is one of our principal motivations behind studying Carroll CFTs, let us mention a few more important applications. 

\medskip

As we have mentioned before, another null manifold of interest is the event horizon of a generic black hole. Indeed, BMS (or ``BMS-like") symmetries  have been found recently on these event horizons \cite{Hawking:2016msc,Hawking:2016sgy,Donnay:2015abr,Afshar:2016wfy,Penna:2017bdn}. Geometrically, Carrollian structures have also been shown to emerge on the event horizons \cite{Donnay:2019jiz}. Naturally, it seems conceivable that a better understanding of the dynamics of black hole event horizons could be achieved by studying BMS invariant field theories living on the event horizons as putative duals. Some of the recent success in this direction involves the computation of black hole entropy using the BMS-Cardy formula \cite{Bagchi:2012xr,Carlip:2017xne,Carlip:2019dbu}.

\medskip

The discussion in the previous paragraphs involved null manifolds mostly in holographic scenarios, where the BMS group is a global symmetry. Carrollian structures also appear in a different context, where the BMS symmetries are gauge symmetries, \emph{viz} the tensionless or null strings \cite{Schild:1976vq,Isberg:1993av,Bagchi:2013bga,Bagchi:2015nca}. A tensionless string, as the name suggests, is obtained by taking the tensionless limit $T\rightarrow 0$ on the tensile string. This is equivalent to the ultra-relativistic limit taking the speed of light on the worldsheet to zero. In this limit, the worldsheet of the string becomes null endowed with a degenerate metric, and acquires a Carrollian structure. On the null worldsheet, the diffeomorphisms are gauge symmetries. For the tensile string, gauge-fixing the worldsheet diffeomorphisms by choosing the conformal gauge results in two copies of the Virasoro algebra as the residual symmetries. Analogously, gauge-fixing of the diffeomorphisms in an equivalent gauge on the null worldsheet results in the BMS$_3$ algebra as the residual symmetries. Thus, like $2d$ CFTs for tensile strings, BMS$_3$ invariant CFTs are inevitable for studying, both classical and quantum, dynamics of tensionless strings. Some recent advances in quantizing the tensionless strings include \cite{Bagchi:2019cay,Bagchi:2020fpr,Bagchi:2021rfw}. 

\medskip

Interestingly, in connection with the discussion of Carroll symmetries on black hole event horizons, it has been recently shown \cite{Bagchi:2020ats, Bagchi:2021ban} that strings moving near the horizon of a black hole spacetime become effectively tensionless. The Carroll structure in the spacetime induces a Carroll structure on the worldsheet as the string hits the horizon of the black hole.

\medskip

From the discussion in the previous paragraphs, we see the essential roles played by Carrollian structures and BMS symmetries in various physical systems of importance. It is, then, evident that a crucial ingredient in advancing these developments further is the systematic study of field theories on null surfaces with Carrollian and BMS symmetries.
%The journey on this avenue has already been started, \eg\ \cite{}.
Indeed there has already been some promising progress in this direction. Carrollian field theories for gauge fields with both Carrollian and conformal Carrollian symmetries have been analyzed in \cite{Duval:2014uoa,Bagchi:2019xfx,Bagchi:2019clu}. Carrollian field theories were also analyzed in the context of fluid dynamics \cite{Ciambelli:2018wre}, dark energy and inflation \cite{deBoer:2021jej}. Some other recent developments include \cite{Basu:2018dub, Henneaux:2021yzg,Hao:2021urq,Chen:2021xkw, Bidussi:2021nmp}. One notices that most of these studies are primarily based on taking the UR limit of relativistic theories at the level of Lagrangian or Hamiltonian, or equations of motions. However we should also look for intrinsic constructions of Carroll covariant field theories, with the belief that they may provide us with novel insights and widespread applicability for these field theories. More importantly one can expect that the space of Carrollian field theories, generically, is larger than those obtained from the UR limit of relativistic theories. Then thoroughly exploring intrinsic descriptions of Carrollian field theories becomes necessary. Some progress towards this has been made: Carroll covariant theories for scalar fields have been investigated in \cite{Ciambelli:2018ojf,Gupta:2020dtl}. The current work aims to provide a self-consistent and explicit realisation of these theories in $2d$.

\subsection{What is in this paper?}

Our primary interest, in this paper, is studying scalar field theories on two dimensional null surfaces. As mentioned before, we want the field theory degrees of freedom to not leave the null surface. This leads us to restricting to theories of massless fields on the null surfaces, which naturally have conformal invariance. Thus we focus on massless field theories on $2d$ Carroll spacetimes, which have conformal Carroll symmetries and therefore are invariant under the BMS$_3$ symmetry algebra. In particular, we focus on classical aspects of these massless scalar field theories in $2d$ and leave description of quantum theories for future work.

\medskip

In a nutshell, we find that there are {\em three inequivalent} Carrollian scalar field theory actions in $d=2$. One of them is the well-known ``time-like" action, studied extensively as we describe below. One is a ``space-like" action, which has also appeared in the literature before. We also find a third and completely new action, which we call the ``mixed" action. In what follows, we will review the time-like action and uncover details of the mixed and the space-like action in detail. We will treat the theory of the mixed action as a field theory, with a fixed background, while our explorations of the space-like action would be in the context of a string theory, where the background is dynamical. The main source of subtlety of the paper is the fact that fixing the background in the second order formulation does not fix all vielbeins in the frame formulation. This leads to various intricacies.

\medskip

We begin this paper by introducing necessary ingredients of the Carroll geometry in sec.~\ref{sec:Carroll-geometry}. For Carroll geometry, the metric is replaced by the data: the degenerate metric $h_{\mu\nu}$ and the no-where vanishing vector field $\tau^{\mu}$ that gives the (null) time direction \cite{Henneaux:1979vn,Duval:2014uoa}. These degenerate Carroll geometries are generally described by two formulations in the literature. The frame formulation of \cite{Hartong:2015xda,Hartong:2015usd,Bergshoeff_2017} in terms of vielbeins is based on local Carroll symmetries in the tangent space along with general diffeomorphisms in the spacetime. The formulation of \cite{Ciambelli:2019lap,Ciambelli:2018wre} is based on Carroll diffeomorphisms in the spacetime. In this paper, we are interested in a formulation which is a composite of the preceding two formulations in a specific way, \ie\ a formulation having both local Carroll symmetries in the tangent space and Carroll diffeomorphisms in the spacetime. In sec.~\ref{sec:Carroll-geometry}, we discuss relevant features of such a composite formulation for Carroll geometry.

\medskip

In the composite formulation, taking contractions of the metric fields with the derivative of the scalar field $\partial_{\mu}\Phi$, we can construct two Carroll covariant second order actions. Contracting with $\tau^{\mu}$ gives the ``timelike" action. This is a well known action and has been studied quite extensively in the literature. This action also represents a single scalar field version of the tensionless string action \cite{Schild:1976vq,Isberg:1993av}. From a field theoretic perspective, this action arises from the ultrarelativistic limit of the free Klein-Gordon action for free scalar field on Minkowski spacetime, see \eg\ \cite{Bergshoeff_2017} for the higher dimensional version. The other action is obtained by contracting $\partial_{\mu}\Phi$ with $h^{\mu\nu}$, the projective inverse of $h_{\mu\nu}$. We call this as the ``spacelike" action following \cite{Ciambelli:2018ojf,Gupta:2020dtl} where a higher dimensional version was introduced in the formulation of \cite{Ciambelli:2018wre,Ciambelli:2019lap} of Carroll geometry.

\medskip

Interestingly, being in $2$-dimensions and using zweibeins in the composite formulation to describe the Carroll geometry leads to another novel Carroll covariant action for the scalar field in $2$-dimensions. Owing to the absence of rotations in $2$-dimensions, we can use the (spatial) zweibein $e^{\mu}$ to contract with $\partial_{\mu}\Phi$. We call the resulting action obtained by using $\tau^{\mu}e^{\nu}$ for contraction with the scalar field as the ``mixed-derivative" action. This looks like the time-space flipped version of the Floreanini-Jackiw (FJ) action \cite{Floreanini:1987as}. We will have more to say about this in Sec \ref{sec:mixed}. 

\medskip

In the composite formulation of the Carroll geometry, the three actions for the scalar field are Carroll covariant and also BMS-Weyl invariant. For all these three actions, we further observe that fixing the Carroll geometry to be torsionless flat Carroll spacetime, the residual symmetries of these gauge-fixed actions are BMS$_3$ symmetries. A crucial point to note here is that owing to the degeneracy of the metric, after choosing the geometry to be torsionless strong flat Carroll spacetime, a component of the zweibein, \ie\ $e^t$ still remains arbitrary. The arbitrariness basically corresponds to the transformation under local Carroll boost, as we will see in sec.~\ref{sec:flat-Carroll}. Being an arbitrary function on the spacetime, $e^t$ also transforms under spacetime conformal Carroll transformations. These transformation properties of $e^t$ play an essential role in ensuring the offshell BMS$_3$ invariance of the actions. A detailed construction of these actions and the symmetry analysis is given in sec.~\ref{mainsec}.

\medskip

In sec.~\ref{sec:timelike}, we review the well known timelike action. We briefly discuss some classical aspects of this action from both the (null) string theory perspective and the field theory perspective. For the spacelike and the mixed-derivative action, the explicit presence of the unspecified vielbein $e^t$ brings non-trivialities in the analysis. In particular, in sec.~\ref{sec:flat-Carroll} we see that from a purely geometric perspective, \ie\ without coupling to any matter the arbitrariness of $e^t_1$ (\ie\ transformations under local Carroll symmetries as well as spacetime Carroll isometries) doesn't affect the geometric quantities of the flat Carroll spacetime, \eg\ the curvatures are still vanishing. However when we couple matter (the scalar field in our case) to the flat Carroll spacetime, we see that $e^t_1$ plays different roles depending on how the matter is coupled. This leads us to treating the spacelike theory and the mixed-derivative theory differently. In sec.~\ref{sec:spacelike}, we find that a string theoretic description is aptly suited for analyzing the spacelike action. We touch upon its residual symmetries and mode expansions. Sec.~\ref{sec:mixed} is the part of our paper with some novel  and unique results, where we analyze the mixed-derivative theory from a field theoretic perspective on a fixed flat Carroll spacetime.  In particular, we study the symmetries, find mode expansions and see an emergence of one copy of Virasoro algebra onshell. In this case, we also see that offshell, we can potentially treat $e^t_1$ as a gauge field as discussed in sec.~\ref{sec:Sm-gauge-symm}. In sec.~\ref{sec:ilst}, we revisit the single scalar field version of the null string in Isberg-Lindstrom-Sundborg-Theodoridis (ILST) formalism \cite{Isberg:1993av}. In particular, we discuss various gauge choices and comment on comparisons with our actions. Appendix \ref{ApA} has some details on comparing the ILST formalism and the (Carroll) geometric formulation of the null string. Appendix \ref{ApB} discusses the parameter space of ILST Lagrange multipliers and various particular forms of these actions in connection to the current work.

\bigskip \bigskip

%%%%%%%%%%%%%%%%%%%%%%%%%%%%%%%%%%%%%%%%%
%%%%%%%% SEC: CARROLL GEOMETRY %%%%%%%%%%

\section{Carroll spacetime in $(d+1)$ dimesions}\label{sec:Carroll-geometry}
As we discussed in the introduction, considering Carrollian theories is equivalent to studying theories on null hypersurfaces characterized by Carroll structures. A Carroll manifold \cite{Henneaux:1979vn,Duval:2014uoa,Duval:2014uva,Duval:2014lpa} in $(d+1)$ dimensions is mostly defined by a degenerate spacetime metric $h_{\mu\nu}$, whose kernel is generated by a no-where vanishing vector field $\zeta^{\mu}$ and a symmetric affine connection $\Gamma^{\rho}_{(\mu\nu)}$. More generally Carroll spacetimes can have torsion \cite{Hartong:2015xda} and are described by a torsion-full connection $\Gamma^{\rho}_{\mu\nu}$. In this paper, we focus on a composite formulation of Carroll geometries based on both local Carroll symmetries in the tangent space and Carroll diffeomorphisms in the spacetime. In this section, we discuss necessary ingredients for the description of Carroll geometries in the composite formulation.
	
\medskip

In the composite formulation, we describe the Carroll spacetime in an adapted coordinate system by
%In an adapted coordinate system it is described by
\begin{equation}\label{curved-Carroll-metric-d}
	ds^2=h_{\mu\nu}dx^{\mu}dx^{\nu} = g_{ij}(t,x^k)dx^i dx^j, \qquad \zeta = \zeta^t(t,x^k)\frac{\partial}{\partial t},
\end{equation}
where $g_{ij}$ is a non-degenerate metric on the $d$-dimensional, exclusively spatial manifold. Here $\mu,\nu$ are spacetime indices denoting the coordinates $(t,x^i)$, where $t$ is a timelike coordinate and $x^i$ are spacelike coordinates $(i=1,...,d)$. In general these are the ingredients that replace the Riemannian structure associated to non-degenerate spacetimes. The Carroll structures \eqref{curved-Carroll-metric-d} are invariant under Carroll diffeomorphisms, where time and space coordinates change in different manners, \ie\
\begin{equation}\label{Carroll-diffeomorphisms}
	t' = t'(t,x^j), \quad x'^i = x'^i(x^j).
\end{equation}

\medskip

We recall that in the frame formulation of Carrollian geometries \cite{Hartong:2015xda,Hartong:2015usd,Bergshoeff_2017}, a $(d+1)$ dimensional Carrollian manifold is described by the vielbeins $\tau_{\mu}(t,x^i)$ and $e^a_{\mu}(t,x^i)$, corresponding to time and space translations. Their projective inverses $\tau^{\mu}(t,x^i)$ and $e^{\mu}_a(t,x^i)$ are defined through the relations
\begin{equation}
	\tau_{\mu}\tau^{\mu} = 1, \quad e^a_{\mu}e^{\mu}_b = \delta^a_b, \quad \tau_{\mu}e^{\mu}_a = 0, \quad e^a_{\mu}\tau^{\mu} = 0, \quad \tau_{\mu}\tau^{\nu} + e^a_{\mu}e^{\nu}_a = \delta^{\nu}_{\mu}. \label{vielbeins-inverse}
\end{equation}
Here $a,b=1,\dots,d$ are spatial tangent space indices.
In addition to the vielbeins, we have the boost spin connection $\Omega_{\mu}^a$ and the rotation spin connection $\Omega_{\mu}^{ab}$ in the tangent space as independent variables. As in the case of Riemannian manifolds, we can determine these tangent space connections in terms of vielbeins by imposing curvature constraints, \ie\ vanishing of the curvatures for local time and space translations. However owing to the degenerate nature, not all components of the spin connections are determined in terms of the vielbeins \cite{Bergshoeff_2017}.

\medskip

Alternately, we can realize the Carroll algebra on a smaller set of fields, \ie\ the vielbeins and an independent vector field, where the spin connections are determined in terms of these fields \cite{Hartong:2015xda}. In any case, the degenerate nature of the spacetime explicitly leads to some independent fields in the tangent space in addition to the vielbeins. We will not go into these details as we won't be needing them for our investigations in this paper. We refer the interested  reader to \cite{Bergshoeff_2017,Hartong:2015xda} for further reading.

\medskip

We also recall that in the frame formulation, the vielbeins transform covariantly under general diffeomorphisms in the spacetime. However, as we will see below shortly, in the composite formulation, the vielbeins get restricted to transform covariantly under Carroll diffeomorphisms only.

\medskip

In the composite formulation, we relate the manifestly Carroll diffeomorphism invariant metric variables in \eqref{curved-Carroll-metric-d} to the vielbeins in the following way:
\begin{eqnarray}
	\zeta^{\mu} = \tau^{\mu}\quad &\Rightarrow& \quad \zeta^t = \tau^t,\  \tau^i = 0, \nonumber \\
	h_{\mu\nu} = e^a_{\mu}e^b_{\nu}\delta_{ab} \quad &\Rightarrow& \quad g_{ij} = e^a_{i}e^b_{j}\delta_{ab},\  e^a_t = 0. \label{vielbeins-metric-relation}
\end{eqnarray}
Solving the vielbein postulates
\begin{equation}
	\partial_{\mu}\tau_{\nu} - \Gamma^{\rho}_{\mu\nu}\tau_{\rho} - \Omega^a_{\mu}e^b_{\nu}\delta_{ab} = 0, \quad \partial_{\mu}e^a_{\nu} - \Gamma^{\rho}_{\mu\nu}e^a_{\rho} - \Omega^{ab}_{\mu}e^c_{\nu}\delta_{bc} = 0,
\end{equation}
we get the spacetime connection in terms of the boost spin connection $\Omega_{\mu}^a$ and the rotation spin connection $\Omega_{\mu}^{ab}$ as
\begin{equation}
	\Gamma^{\rho}_{\mu\nu} = \tau^{\rho}\partial_{\mu}\tau_{\nu} - \tau^{\rho}\Omega_{\mu}^a e_{\nu}^b\delta_{ab} + e^{\rho}_a\partial_{\mu}e^a_{\nu} -e^{\rho}_a\Omega_{\mu}^{ab}e^c_{\nu}\delta_{bc}.\label{connections-relation}
\end{equation}

\medskip

We can see from \eqref{vielbeins-metric-relation} that relating vielbeins to manifestly Carroll diffeomorphism invariant metric variables in \eqref{curved-Carroll-metric-d} restricts the vielbeins to transform covariantly under Carroll diffeomorphisms only. Under these Carroll-diffeomorphisms \eqref{Carroll-diffeomorphisms} (compactly written as $x'^{\mu} = x'^{\mu}(x^{\nu})$), the vielbeins and their (projective) inverses transform as
\begin{equation}\label{Carroll-diffeos-d}
	\tau'^{\mu} = \frac{\partial x'^{\mu}}{\partial x^{\nu}}\tau^{\nu}, \quad e'^{\mu}_a = \frac{\partial x'^{\mu}}{\partial x^{\nu}}e^{\nu}_a, \quad \tau'_{\mu} = \frac{\partial x^{\nu}}{\partial x'^{\mu}}\tau_{\nu}, \quad e'^a_{\mu} = \frac{\partial x^{\nu}}{\partial x'^{\mu}}e^a_{\nu}.
\end{equation}
Moreover, a Carroll-Weyl transformation \cite{Ciambelli:2019lap}, the cousin of regular Weyl transformations, albeit on null hypersurfaces, is defined as
\begin{equation}\label{Carroll-Weyl-d}
	\tau_{\mu} \rightarrow e^{z\Omega(t,x)}\tau_{\mu}, \quad h_{\mu\nu} \rightarrow e^{2\Omega(t,x)}h_{\mu\nu}.
\end{equation}
As one can note, this implies that the metric components associated to the temporal part and the spatial part scale differently under a Weyl transformation, which is a crucial notion in Carrollian geometry. The isomorphism between Carrollian Conformal Algebras (CCA) in $d$ dimensions and BMS algebras in $d+1$ dimensions requires the exponent value fixed at $z=1$ \cite{Duval:2014uva,Duval:2014lpa}. 
Hence for $z=1$, this Carroll-Weyl transformation gives rise to the BMS-Weyl transformations of the vielbiens,
\begin{equation}\label{BMS-Weyl-d}
	\tau_{\mu} \rightarrow e^{\Omega(t,x)}\tau_{\mu},\quad e^a_{\mu} \rightarrow e^{\Omega(t,x)}e^a_{\mu},\quad \tau^{\mu} \rightarrow e^{-\Omega(t,x)}\tau^{\mu},\quad e^{\mu}_a \rightarrow e^{-\Omega(t,x)}e^{\mu}_a,
\end{equation}
signalling the generation of conformal isometries of the Carrollian manifolds we described earlier. 

\medskip

So far, we discussed Carroll diffeomorphisms on the spacetime. Now we turn to the second fundamental ingredient of the composite formulation, \ie\ the local Carroll symmetries in the tangent space. Under a set of (infinitesimal) local Carroll boosts and spatial rotations in the tangent space, parametrized by $\lambda_a$ and $\lambda^a_{\ b}$ respectively, the vielbeins transform as
\begin{equation}
	\delta \tau^{\mu} = 0, \quad \delta e^{\mu}_a = -\tau^{\mu}\lambda_a + \lambda_a^{\ b}e^{\mu}_b, \quad \delta \tau_{\mu} = e^a_{\mu}\lambda_a, \quad \delta e^a_{\mu} = \lambda^a_{\ b}e^b_{\mu} \label{vielbeins-local-transformations}.
\end{equation}
Using $h_{\mu\nu} = e^a_{\mu}e^b_{\nu}\delta_{ab}$, $h^{\mu\nu} =  e^{\mu}_a e^{\nu}_b \delta^{ab}$, the transformation of the degenerate spacetime metric $h_{\mu\nu}$ and the projective inverse $h^{\mu\nu}$ follows from \eqref{vielbeins-local-transformations}:
\begin{equation}\label{metric-local-transformations}
	\delta h_{\mu\nu} = 0, \quad \delta h^{\mu\nu} = -(\tau^{\mu}e^{\nu}_a + \tau^{\nu}e^{\mu}_a)\lambda^a.
\end{equation}

\medskip

We now see how these local transformations of vielbeins are obtained, and contrast these in the frame formalism with those in the composite formalism. We begin by recalling that the ordinary derivatives $\partial_{\mu}$ constitute a basis in the tangent space.

\medskip

In the \emph{frame formalism} in \cite{Hartong:2015xda,Bergshoeff_2017}, we do a change of basis in the tangent space as
\begin{equation}
	e_0 = \tau^{\mu}\partial_{\mu}, \quad e_a = e^{\mu}_a\partial_{\mu}, \label{new-basis}
\end{equation}
and likewise $e^0 = \tau_{\mu}dx^{\mu}$, $e^a = e_{\mu}^a dx^{\mu}$ in the co-tangent space. Then under local Carroll transformations in the tangent space (boosts $\Lambda^0_{\ a} = -\lambda_a$ and rotations $\Lambda^a_{\ b}=\delta^a_b + \lambda^a_{\ b}$), the new basis $(e^0,e^a)$, $(e_0,e_a)$ transform like the $t$ and $x^i$ directions, \ie
\begin{equation}
	e'^0 = e^0 -\lambda_a e^a, \quad e'^a = \Lambda^a_{\ b}e^b, \quad e'_0 = e_0, \quad e'_a = (\Lambda^{-1})^{\ b}_a (e_b + \lambda_b e_0).
\end{equation}
Simplifying these equations, using \eqref{new-basis} and $\delta_{\lambda}\partial_{\mu} = 0$ gives the local transformations of vielbeins in \eqref{vielbeins-local-transformations}.

\medskip

In the \emph{composite formalism}, we do a change of basis from $\partial_{\mu}$ to projected directions in the tangent space:
\begin{equation}
	\hat{e}_0 = \hat{\tau}^{\mu}\partial_{\mu}, \quad \hat{e}_a = \hat{e}^{\mu}_a\partial_{\mu}, \label{new-projected-basis}
\end{equation}
and likewise $\hat{e}^0 = \hat{\tau}_{\mu}dx^{\mu}$, $\hat{e}^a = \hat{e}_{\mu}^a dx^{\mu}$. By projected directions, we mean that under local Carroll transformations, the basis $(\hat{e}_0,\hat{e}_a)$, $(\hat{e}^0,\hat{e}^a)$ transform as ``Carrollian" tensors with respect to the base space (of the tangent space):
\begin{equation}
	\hat{e}'^0 = \hat{e}^0, \quad \hat{e}'^a = \Lambda^a_{\ b}\hat{e}^b, \quad \hat{e}'_0 = \hat{e}_0, \quad \hat{e}'_a = (\Lambda^{-1})^{\ b}_a \hat{e}_b, \label{projected-vielbeins-transf}
\end{equation}
\ie\ $\hat{e}^0$ is a scalar and $\hat{e}^a$ is a spatial vector in the tangent space, and $\hat{e}_0$ is a zero-form and $\hat{e}_a$ is a (spatial) one-form in the co-tangent space. Simplifying the transformations \eqref{projected-vielbeins-transf} using \eqref{new-projected-basis}, we get that the projected vielbeins transform as given in \eqref{vielbeins-local-transformations} provided the ordinary derivatives $\partial_{\mu}$ transform as 
\begin{equation}
	\delta_{\lambda}\partial_t = 0, \quad \delta_{\lambda}\partial_i = \lambda_i\partial_t, \label{derivative-local-transf}
\end{equation}
where $\lambda_{\mu} = e^a_{\mu}\lambda_a$ and $\lambda_{\mu}\tau^{\mu} = 0$. These local transformations of the (projected) vielbeins and the ordinary derivatives will be used to show the local Carroll invariance of the actions in sec.~\ref{sec:actions-local-invariance}.

\medskip

We see that \eqref{projected-vielbeins-transf} also admits another solution:
\begin{eqnarray}
	&& \delta_{\lambda}(\partial_{\mu}) = 0, \quad \delta_{\lambda}(dx^{\mu}) = 0; \nonumber \\
	&& \delta_{\lambda}\hat{\tau}^{\mu} = 0, \quad \delta_{\lambda}\hat{\tau}_{\mu} = 0, \quad \delta_{\lambda}\hat{e}^{\mu}_a = -\lambda^b_{\ a}\hat{e}^{\mu}_b, \quad \delta_{\lambda}\hat{e}_{\mu}^a = \lambda^a_{\ b}\hat{e}^b_{\mu}.
\end{eqnarray}
It follows that the degenerate metric $\hat{h}_{\mu\nu} = \hat{e}^a_{\mu}\hat{e}^b_{\nu}\delta_{ab}$ and its projective inverse $\hat{h}^{\mu\nu} = \hat{e}_a^{\mu}\hat{e}_b^{\nu}\delta^{ab}$ transform as
\begin{equation}
	\delta_{\lambda}\hat{h}_{\mu\nu} = 0, \quad \delta_{\lambda}\hat{h}^{\mu\nu} = 0.
\end{equation}
In this paper, we focus on \eqref{derivative-local-transf}, and leave the investigation of this alternate solution to a future work.

\medskip

The reason that allows us to use the projected basis \eqref{new-projected-basis} is the degenerate nature of the Carrollian tangent space. Due to this degenerate nature, we can change the basis in the tangent space from $\partial_{\mu}$ to either the standard basis \eqref{new-basis} or the projected basis \eqref{new-projected-basis}. To see this more explicitly, we note that the metric data of the flat Carroll tangent space looks the same in standard directions as well as projected directions, \ie
\begin{equation}
	\tau^A = (1,0), \quad h_{AB} = \delta_{ab}; \qquad \hat{\tau}^A = (1,0), \quad \hat{h}_{AB} = \delta_{ab}, \label{tangent-space-metric}
\end{equation}
where $A=0,a$. To emphasize again, it is because of \eqref{tangent-space-metric} that we can have the projected basis \eqref{new-projected-basis} and their local transformations \eqref{projected-vielbeins-transf} in the composite formalism.

\medskip

In the rest of the paper, we will use the composite formalism and projected vielbeins. For the convenience of notation, we will drop the `hat' and call `projected vielbeins' as just `vielbeins'.

\subsection{Flat Carroll spacetime in generic dimensions}\label{sec:flat-Carroll}

A strong flat Carroll spacetime \cite{Henneaux:1979vn,Duval:2014uoa,Duval:2014uva,Duval:2014lpa} in $(d+1)$ dimensions is defined in terms of one having vanishing (symmetric) affine connection and particular choice of the spatial metric and vector field,
\begin{equation}\label{flat-Carroll-metric-d}
	ds^2=h_{\mu\nu}dx^{\mu}dx^{\nu} = \delta_{ij}dx^i dx^j, \qquad \zeta = \frac{\partial}{\partial t}, \qquad \Gamma^{\rho}_{(\mu\nu)} = 0.
\end{equation}
The Carroll group compositions are made up of those transformations that  leave this flat connection invariant.
For the flat Carroll spacetime in \eqref{flat-Carroll-metric-d}, using \eqref{vielbeins-metric-relation} and \eqref{vielbeins-inverse}, we can solve for the values of vielbeins as
\begin{eqnarray}\label{zweis}
	&& \tau_{t} = 1, \quad e^a_t = 0, \quad e^a_i = \delta^a_i, \quad \tau_i \sim \text{arbitrary}, \nonumber \\
	&& \tau^t = 1, \quad \tau^i = 0, \quad e^i_a = \delta^i_a, \quad e^t_a = -\delta^i_a\tau_i. \label{flat-Carroll-vielbeins}
\end{eqnarray}
Our focus, in this paper, would be on field theories on a torsionless flat Carroll background. Imposing that the torsion, \ie\ the antisymmetric part of the affine connection vanishes, in addition to the symmetric part, leads to imposing that the full affine connection summarily  vanishes. Then using \eqref{connections-relation} to solve $\Gamma^{\rho}_{\mu\nu} = 0$ for flat vielbeins \eqref{flat-Carroll-vielbeins}, we get the tangent space connections
\begin{equation}
\Omega_{\mu}^{ab} = 0, \quad \Omega_{\mu}^a = \delta^{ai}\partial_{\mu}\tau_i. \label{flat-boost-connection}
\end{equation}
For these expressions of the vielbeins \eqref{flat-Carroll-vielbeins} and the spin connections \eqref{flat-boost-connection} for the torsionless flat Carroll spacetime \eqref{flat-Carroll-metric-d}, the curvatures for the boost spin-connection and the rotation spin-connection also vanish identically:
\begin{eqnarray}
	R^a_{\mu\nu}(C) = 2\partial_{[\mu}\Omega_{\nu]}^a - 2\Omega_{[\mu}^{ab}\Omega_{\nu]}^c\delta_{bc} = 0, \quad R^{ab}_{\mu\nu}(J) = 2\partial_{[\mu}\Omega_{\nu]}^{ab} - 2\Omega_{[\mu}^{ac}\Omega_{\nu]}^{db}\delta_{cd} = 0, \label{local-curvatures}
\end{eqnarray}
where $C_a$ and $J_{ab}$ are generators of local boosts and rotations, respectively, on the tangent space. This ultimately leads to the vanishing of Riemann tensor on the spacetime, and this can be seen in terms of the above curvatures using the vielbein postulates:
\begin{equation}
	\mathcal{R}_{\mu\nu\sigma}^{\quad\ \rho} = \tau^{\rho}e_{\sigma a}R_{\mu\nu}^a(C) - e_{\sigma a}e^{\rho}_b R_{\mu\nu}^{ab}(J) = 0. \label{Carroll-Riemann-tensor}
\end{equation}
Note that, in our definition of the flat Carroll spacetime in terms of fixing the spacetime data \eqref{flat-Carroll-metric-d} with vanishing Riemann tensor \eqref{Carroll-Riemann-tensor}, the spatial components of the vielbein $\tau_i$ remain arbitrary with a non-vanishing boost spin-connection \eqref{flat-boost-connection}. This arbitrariness of $\tau_i$, \ie\ $e^t_a$ is a crucial point here, as it allows us to construct novel actions for fields coupled to the flat Carroll background, as we will elaborate on in later sections.\\

\subsection*{A gauge redundancy in $\tau_i$}
Under local Carroll boost in the tangent space \eqref{vielbeins-local-transformations}, the components of $\tau_{\mu}$ transform as
\begin{equation}
	\delta\tau_t = 0,\  \delta \tau_i = \lambda_i, \quad \Rightarrow \quad \delta\Omega_{\mu}^a = \delta^{ai}\partial_{\mu}\lambda_i, \label{tau_i-gauge-transformation}
\end{equation}
which leaves the boost curvature \eqref{local-curvatures} invariant. We observe that the transformation of the boost spin connection here is a gauge transformation, if we think of $\Omega_{\mu}^a$ as a gauge field in the tangent space.
An instructive exercise here is to start with diffeomorphism and Weyl invariant relativistic field theory on a Randers-Papapetrou type background \cite{Ciambelli:2019lap}:
\begin{equation}
	ds^2 = -c^2(\omega dt - b_i dx^i)^2 + a_{ij}dx^i dx^j = (-c^2 \tau_{\mu}\tau_{\nu} + e^a_{\mu}e^b_{\nu}\delta_{ab})dx^{\mu}dx^{\nu},
\end{equation}
and taking the ultrarelativistic limits ($c\rightarrow 0$) where it produces a Carroll geometry. In this limit, we can explicitly
relate our vielbeins to the metric variables,
\begin{equation}
	\omega = \tau_t, \quad b_i = -\tau_i, \quad a_{ij} = e^a_i e^b_j\delta_{ab}. \label{vielbeins-RP-metric}
\end{equation}

This identifies Carrollian spacetime as a fiber bundle structure endowed with an Ehresmann connection $b_i$, that allows for separate manners of transformation for time and space under Carroll diffeomorphisms. 
We can now clearly identify the gauge redundancy of the Ehresmann connection $b_i$ for the flat Carroll spacetime \eqref{flat-Carroll-metric-d} introduced in  \cite{Ciambelli:2019lap} as the transformation of $\tau_i$ under local Carroll boost in the tangent space \eqref{tau_i-gauge-transformation}.

\subsection{Carroll spacetime in two dimensions}\label{sec:flat-Carroll-2d}
In this work, we will mostly be focussing on scalar fields living on a two dimensional Carrollian manifold, hence a zoom-in into this particular example seems useful here. In conformity with our discussion above for general dimensions, the $2d$ version of the Carrollian geometry is described by the zweibeins $\tau_{\mu}(t,x)$ and $e^1_{\mu}(t,x)$, with their inverses $\tau^{\mu}(t,x)$ and $e^{\mu}_1(t,x)$ defined through the relations
\begin{equation}
	\tau_{\mu}\tau^{\mu} = 1, \quad e^1_{\mu}e^{\mu}_1 = 1, \quad \tau_{\mu}e^{\mu}_1 = 0, \quad e^1_{\mu}\tau^{\mu} = 0, \quad \tau_{\mu}\tau^{\nu} + e^1_{\mu}e^{\nu}_1 = \delta^{\nu}_{\mu}. \label{zweibeins-inverse}
\end{equation}
For our discussion in later sections, we can redefine the timelike zweibeins in these dimensions as $$\tau_{\mu} = e^0_{\mu}, ~\tau^{\mu} = e^{\mu}_0.$$ 
We will be using the above notations interchangeably throughout this work. 
Now we can collectively write the zweibeins and their inverses as $e^{\mu}_A$, ($A=0,1$) in terms of which the above relations can be written in a compact form:
\begin{equation}
	e^A_{\mu}e^{\mu}_B = \delta^A_B, \qquad e^A_{\mu}e^{\nu}_A = \delta^{\nu}_{\mu}.
\end{equation}
We further introduce antisymmetric symbols $\epsilon^{\mu\nu}$ and $\epsilon^{AB}$ defined as $\epsilon^{tx}=1=-\epsilon_{tx}$ and $\epsilon^{01}=1=-\epsilon_{01}$, which satisfy $\epsilon_{\mu\rho}\epsilon^{\rho\nu}=\delta_{\mu}^{\nu}$ and $\epsilon_{AC}\epsilon^{CB}=\delta_A^B$. Using these antisymmetric symbols, we can write the determinants $e=\det(e^A_{\mu})$, $\frac{1}{e}=\det(e^{\mu}_A)$ and the inverse zweibeins in a simple form,
\begin{equation}
	e=\frac{1}{2}\epsilon_{AB}e^A_{\mu}e^B_{\nu}\epsilon^{\nu\mu}, \qquad \frac{1}{e} = \frac{1}{2}\epsilon^{AB}e^{\mu}_A e^{\nu}_B\epsilon_{\nu\mu}, \qquad e^{\mu}_A = \frac{1}{e}\epsilon^{\mu\nu}e^B_{\nu}\epsilon_{BA}.
\end{equation}

\medskip

We also need to spell out the transformation of the zweibeins under spacetime Carroll-diffeomorphisms $x^{\mu}\rightarrow x'^{\mu}(x^{\nu})$, \ie\ $t'=t'(t,x)$ and $x'=x'(x)$ in this case, which can be written in our compact notation as
\begin{equation}\label{Carroll-diffeos}
	e'^{\mu}_A = \frac{\partial x'^{\mu}}{\partial x^{\nu}}e^{\nu}_A, \quad e'^A_{\mu} = \frac{\partial x^{\nu}}{\partial x'^{\mu}}e^A_{\nu}.
\end{equation}
As before a set of Carroll-Weyl transformations \cite{Ciambelli:2019lap} are defined as
\begin{equation}\label{Carroll-Weyl}
	e^0_{\mu} \rightarrow e^{z\Omega(t,x)}e^0_{\mu}, \quad h_{\mu\nu} \rightarrow e^{2\Omega(t,x)}h_{\mu\nu},
\end{equation}
 with $h_{\mu\nu} = e^1_{\mu}e^1_{\nu}$. As mentioned earlier, this Carroll-Weyl transformation gives the BMS-Weyl transformation when $z=1$.

\medskip
\noindent In $2d$ there is only one local Carroll boost and no spatial rotations in the tangent space. Hence under an infinitesimal local Carroll boost parametrized by $\lambda_1$, the zweibeins transform as
\begin{equation}
	\delta e^{\mu}_0 = 0, \quad \delta e^{\mu}_1 = -e^{\mu}_0\lambda_1, \quad \delta e^0_{\mu} = e^1_{\mu}\lambda_1, \quad \delta e^1_{\mu} = 0, \label{zweibeins-local-transformations}
\end{equation}
and the degenerate spacetime metric transforms as expected
\begin{equation}
	\delta h_{\mu\nu} = 0, \quad \delta h^{\mu\nu} = -(\tau^{\mu}e^{\nu}_1 + \tau^{\nu}e^{\mu}_1)\lambda^1. \label{2d-h-local-transformations}
\end{equation}

\medskip

\subsection{Flat Carroll manifold in $2d$}\label{sec:2d-flat-Carroll}
The definition of a torsionless strong flat Carrollian spacetime in $2d$ follows our previous discussion, but simplifies considerably as only one spatial direction is involved
\begin{equation}\label{flat-Carroll-metric}
	ds^2=h_{\mu\nu}dx^{\mu}dx^{\nu} = dx^2, \qquad \zeta = \frac{\partial}{\partial t}, \qquad \Gamma^{\rho}_{\mu\nu} = 0.
\end{equation}
Using the relation \eqref{vielbeins-metric-relation} (for the index $a=1$) between metric variables in the second order formalism and the zweibeins, we can identify the values of the flat zweibeins as
\begin{equation}
	h_{\mu\nu} = e^1_{\mu}e^1_{\nu}, \quad \zeta^{\mu} = e^{\mu}_0=\tau^\mu
\end{equation}
and the flat connection in $(1+1)d$ reads
\begin{equation}
	\Gamma^{\rho}_{\mu\nu} = e^{\rho}_0\partial_{\mu}e^0_{\nu} - e^{\rho}_0\Omega_{\mu}^1e_{\nu}^1 + e^{\rho}_1\partial_{\mu}e^1_{\nu}, \label{2d-affine-connection}
\end{equation}
where $\Omega_{\mu}^1$ is the boost spin-connection in the tangent space. Then for \eqref{flat-Carroll-metric} we get the values of all zweibeins as
\begin{eqnarray}
	&& e^0_{t} = 1, \quad e^1_t = 0, \quad e^1_x = 1, \quad e^0_x \sim \text{arbitrary}, \nonumber \\
	&& e^t_0 = 1, \quad e^x_0 = 0, \quad e^x_1 = 1, \quad e^t_1 = -e^0_x, \label{flat-Carroll-zweibeins}
\end{eqnarray}
and the dependent boost spin-connection has a form $\Omega_{\mu}^1 = \partial_{\mu}e^0_x$, which echoes (\ref{zweis}), \eqref{flat-boost-connection} for $2d$. One again notes here that gauge fixing the spatial metric and the vector field for a flat Carroll spacetime cannot fully fix the particular vielbein $e^t_1 = -e^0_x$. Although gauge fixing this to a constant will make  the boost spin-connection vanish, we don't elect to choose so at this moment. This arbitrariness in $e^t_1$, as alluded to before, will be a central theme in this work and we will continuously return to this subtlety in the later sections.

\subsubsection*{More on Flat Carroll manifolds}
In the discussion above, the degenerate metric for the flat Carroll manifold in $2d$ was chosen as $h = \text{diag} (0,1)$. One may be tempted to interpret this choice as the most obvious one coming from a Carrollian limit of the Minkowski metric in two dimensions i.e. $\eta = \text{diag} (-c^2,1)$. However, one might argue that there are more possible choices for the degenerate metric which satisfies all tenets of being in the flat Carroll class. Let us discuss then another choice of $h$ that will be important to us later, 

\begin{equation}
	ds^2=h_{\mu\nu}dx^{\mu}dx^{\nu} = (dx^-)^2, \qquad \zeta = 2\frac{\partial}{\partial x^+}, \qquad \Gamma^{\rho}_{\mu\nu} = 0. \label{flat-Carroll-metric-v2}
\end{equation}
Of course the reader may ask, what is the most general choice for $h$ that is compliant with all the Carroll postulates? We'll come back to this question later, for now concentrating solely on the above case. One may also note that this particular degenerate metric is not connected to the Minkowski metric via a Carrollian limit.

\medskip

\noindent For this parametrization of the degenerate metric $h_{\mu\nu}$ and the vector field $\zeta^{\mu}$ in \eqref{flat-Carroll-metric-v2}, using $x^{\pm} = t\pm x$ and \eqref{zweibeins-inverse}, the components of the zweibeins are computed, 
\begin{eqnarray}
	&& \tau^t = 1, \quad \tau^x = 1, \quad \tau_t = 1 - \tau_x, \quad \tau_x \sim \text{arbitrary}, \nonumber \\
	&& e_t = -1, \quad e_x = 1, \quad e^t = -\tau_x, \quad e^x = 1- \t_x.
\end{eqnarray}
The vanishing of the spacetime connection \eqref{2d-affine-connection}, $\Gamma^{\rho}_{\mu\nu} = 0$ gives $\Omega_{\mu} = \partial_{\mu}\tau_x$, given in terms of the arbitrary unfixed vielbein $\t_x$ as before. This form of the boost spin-connection again leads to vanishing of boost curvature $R_{\mu\nu}(C) = 2\partial_{[\mu}\Omega_{\nu]} = 0$, and hence vanishing Riemann tensor $\mathcal{R}_{\mu\nu\sigma}^{\quad\ \rho} = \tau^{\rho}e_{\sigma}R_{\mu\nu}(C) = 0$. All of this makes sure this new choice still belongs to the flat Carroll class.

\medskip

With all the components of the geometry at our disposal, we will start looking at field theories living on this $2d$ flat Carroll spacetime. In what follows, we will mostly be working with the standard flat Carroll choice of $h = \text{diag} (0,1)$, unless otherwise specified. 

\medskip

\section{BMS invariant actions for a $2d$ scalar field}\label{mainsec}
$2d$ scalar field theories invariant under the BMS$_3$ algebra \refb{bms3} have previously been studied at length in connection to null string theories \cite{Bagchi:2013bga,Bagchi:2015nca, Bagchi:2020fpr} and more recently as field theory in  \cite{Hao:2021urq}. These in general rely on null or Carrollian limits taken from conformal scalar field and/or conformal gauge fixed string theories. On general grounds, it is expected that considering (quantum) field theories on an inherently Carroll manifold would give rise to a wider class of theories than those appearing in the Carroll limit. With the Carroll covariant formalism at our behest, we can now make this more robust by exploiting the isomorphism of conformal Carroll groups and BMS groups. We will focus on scalar field theories. 

\medskip

To start with, we construct independent classes of Carroll covariant actions for a free scalar field $\Phi(x,t)$ on a 2d Carroll spacetime that are manifestly invariant under Carroll-diffeomorphisms \eqref{Carroll-diffeos} and BMS-Weyl transformations \eqref{Carroll-Weyl}. Using the composite formalism discussed in the last section, we find {\em{three distinct ones}}:
\begin{subequations}\label{threeactions}
\begin{eqnarray}
&& S_{00} = \int dtdx\, e\, e^{\mu}_0e^{\nu}_0\,\partial_{\mu}\Phi\partial_{\nu}\Phi, \\
&& S_{01} = \int dtdx\, e\, e^{\mu}_0e^{\nu}_1\,\partial_{\mu}\Phi\partial_{\nu}\Phi, \\
&& S_{11} = \int dtdx\, e\, e^{\mu}_1e^{\nu}_1\,\partial_{\mu}\Phi\partial_{\nu}\Phi.
\end{eqnarray}
\end{subequations}
The Carroll diffeomorphism invariance of each of the actions can be seen from a straight-forward calculation. To see the BMS-Weyl invariance, we note that the scalar field has a BMS-Weyl weight zero, \emph{i.e.} it doesn't transform under the BMS-Weyl transformation. Introducing arbitrary constants \footnote{It is fine to do so in two dimensions, however there may be more than what meets the eye for higher dimensional constructions. See sec.~\ref{sec:actions-local-invariance} for a discussion.} $W^{AB}$ that are symmetric in $A,B$ ($A,B,\dots = 0,1$), we can collectively write the three Carroll covariant actions into a single Carroll covariant action:
\begin{equation}
	S = W^{00}S_{00} + 2W^{01}S_{01} + W^{11}S_{11} = W^{AB}\int dtdx\, e\, e^{\mu}_A e^{\nu}_B\,\partial_{\mu}\Phi\partial_{\nu}\Phi. \label{total-action}
\end{equation}
We would like to emphasize that $W^{AB}$ are just arbitrary constants introduced for our convenience to write the unified Carroll covariant action \eqref{total-action}. More precisely, $W^{AB}$ should not be thought of as the flat metric on the tangent space, since the tangent space is degenerate as well and the correct (flat) geometric data describing the tangent space is $\tau^{A}=(1,0)$ and the spatial metric $h_{AB} = \text{diag}(0,1)$. In the rest of the section, we will explicitly discuss the symmetry invariance for all these three component actions. 

\subsection{BMS invariance on flat Carroll background}
Defined on the flat Carroll background \eqref{flat-Carroll-zweibeins}, we will call the three actions we encountered in (\ref{threeactions}) ``timelike", ``mixed-derivative" and ``spacelike" actions respectively,
\begin{eqnarray}
	&& S_{00}\ \rightarrow \ S_t = \int dtdx (\partial_t\Phi)^2, \nonumber \\
	&& S_{01}\ \rightarrow \ S_{m} = \int dtdx \big( \partial_x\Phi \partial_t\Phi + e^t_1(\partial_t\Phi)^2 \big), \nonumber \\
	&& S_{11}\ \rightarrow \ S_{sp} = \int dtdx ( \partial_x\Phi + e^t_1 \partial_t\Phi)^2. \label{gauge-fixed-actions}
\end{eqnarray}
The names of course reflect their spacetime structures. Once we choose the background geometry to be flat Carroll spacetime, these actions \eqref{gauge-fixed-actions} are no longer invariant under Carroll-diffeomorphisms and BMS-Weyl transformations, as expected. However they are invariant under a residual set of transformations. To find these residual symmetries, we gauge fix the degenerate metric $h_{\mu\nu}$ and the vector field $\zeta^{\mu}$ to their flat spacetime values \eqref{flat-Carroll-metric} such that they are invariant under a combined action of a diffeomorphism and a BMS-Weyl transformation. 

\medskip

An infinitesimal diffeomorphism is given by $$x^{\mu}\rightarrow x^{\mu} + \xi^{\mu},$$ and we choose the BMS-Weyl factor $\Omega(t,x)$ be small. Then the gauge-fixing of $h_{\mu\nu}$ and $\zeta^{\mu}$ to flat spacetime values leads to the conformal Killing equations
\begin{equation}
	\mathcal{L}_{\xi} h_{\mu\nu} = 2\Omega h_{\mu\nu}, \qquad \mathcal{L}_{\xi}\zeta^{\mu} = -\Omega \zeta^{\mu},
\end{equation}
where $\mathcal{L}$ is the Lie derivative of these fields under the diffeomorphism. Solving these equations for the flat Carroll spacetime \eqref{flat-Carroll-metric}, we can constrain the components of the vector field $\xi$,
\begin{equation}\label{cct}
	\partial_t\xi^t = \partial_x\xi^x, \quad \partial_t\xi^x = 0, \quad \Omega = \partial_x\xi^x.
\end{equation}
We can recognize these as BMS$_3$ transformations, which give rise to the solutions for the Killing fields
\be{bmstrns}
\xi^t = f'(x)t+g(x),~~\xi^x = f(x),
\ee
where $f'(x) = \partial_x f$. Under these conformal Carroll transformations, the inverse zweibeins transform as
\begin{equation}
	e'^{\mu}_A(x') e^{\Omega} = \frac{\partial x'^{\mu}}{\partial x^{\nu}}e^{\nu}_A(x).
\end{equation}
From this we can now explicitly check that $e^t_0$, $e^x_1$ and $e^x_0$ are all invariant under these conformal Carroll transformations. Thus we see that gauge fixing $h_{\mu\nu}$ and $\zeta^{\mu}$ amounts to gauge-fixing $e^t_0$, $e^x_1$ and $e^x_0$, while leaving $e^t_1$ arbitrary, as found earlier. Starting from the transformation law for the unfixed zweibein $e^t_1$ in the form
\begin{equation}
	e'^t_1(x')  = e^{-\Omega(x)} \frac{\partial t'}{\partial x^{\nu}}e^{\nu}_1(x),
\end{equation}
we get the corresponding transformation law for small $\Omega$ and $\xi$ 
\begin{eqnarray}
	e'^t_1(x') = e^t_1 + e^{\nu}_1\partial_{\nu}\xi^t -\Omega e^t_1 = e^t_1 + \partial_x\xi^t. \label{et1-transformation}
\end{eqnarray}
Using this result, we can go ahead to find the conformal Carroll transformation \eqref{cct} of derivatives of the scalar field:
\begin{eqnarray}
	&& \partial'_{\mu}\Phi'(x') = \partial_{\mu}\Phi(x) - \partial_{\mu}\xi^{\nu}\partial_{\nu}\Phi(x), \nonumber \\
	i.e.\quad &&  \partial_{t'}\Phi'(x') = \partial_t\Phi (1-\partial_t\xi^t), \qquad \partial_{x'}\Phi'(x') = \partial_x\Phi (1-\partial_x\xi^x) - \partial_x\xi^t \partial_t\Phi. \label{delPhi-transformation}
\end{eqnarray}

\medskip

\noindent Using \eqref{et1-transformation} and \eqref{delPhi-transformation}, we now show that the each of the action in \eqref{gauge-fixed-actions} is invariant under BMS$_3$ \ie\  the set of conformal Carroll transformations \eqref{cct}. Below we give detailed expressions below for all three classes of actions. 

\medskip

\noindent \underline{\textbf{(1) Timelike action}\,:}
\begin{eqnarray}
	S'_t &=& \int dt'dx' (\partial_{t'}\Phi')^2 \nonumber \\
	&=& \int dtdx (1+2\partial_t\xi^t) (1-\partial_t\xi^t)^2(\partial_t\Phi)^2 \nonumber \\
	&=& \int dtdx (\partial_t\Phi)^2 = S_t.
\end{eqnarray}

\noindent\underline{\textbf{(2) Mixed-derivative action}\,:}
\begin{eqnarray}
	S'_{m} &=& \int dt'dx' \Big( \partial_{x'}\Phi' \partial_{t'}\Phi' + e'^t_1(\partial_{t'}\Phi')^2 \Big) \nonumber \\
	&=& \int dtdx (1+2\partial_t\xi^t)\Big( (\partial_x\Phi(1-\partial_x\xi^x)-\partial_x\xi^t\partial_t\Phi)(1-\partial_t\xi^t)\partial_t\Phi \nonumber \\
	&& \qquad + (e^t_1 + \partial_x\xi^t)(1-2\partial_t\xi^t)(\partial_t\Phi)^2 \Big) \nonumber \\
	&=& \int dtdx \Big( \partial_x\Phi \partial_t\Phi + e^t_1(\partial_t\Phi)^2 \Big) = S_{m}.
\end{eqnarray}

\noindent \underline{\textbf{(3) Spacelike action}\,:}
\begin{eqnarray}
	S'_{sp} &=& \int dt'dx' \Big( (\partial_{x'}\Phi')^2 + 2 e'^t_1 \partial_{x'}\Phi'\partial_{t'}\Phi' +  (e'^t_1)^2(\partial_{t'}\Phi')^2 \Big) \nonumber \\
	&=& \int dtdx (1+2\partial_t\xi^t)\Big((\partial_x\Phi)^2(1-2\partial_x\xi^x) - 2\partial_x\xi^t\partial_x\Phi\partial_t\Phi \nonumber \\
	&& \qquad +2(e^t_1 + \partial_x\xi^t)(\partial_x\Phi(1-\partial_x\xi^x)-\partial_x\xi^t\partial_t\Phi)(1-\partial_t\xi^t)\partial_t\Phi \nonumber \\
	&& \qquad + ((e^t_1)^2 + 2e^t_1\partial_x\xi^t)(1-2\partial_t\xi^t)(\partial_t\Phi)^2 \Big)  \nonumber \\
	&=& \int dtdx \Big( (\partial_x\Phi)^2 + 2 e^t_1 \partial_x\Phi\partial_t\Phi +  (e^t_1)^2(\partial_t\Phi)^2 \Big) = S_{sp}.
\end{eqnarray}

With the explicit invariance under BMS$_3$ transformations shown, some comments are in order here. As the well versed reader can spot, the ``timelike'' action in our notion is indeed the single scalar field version of the null string action \cite{Isberg:1993av}, and also the action discussed in  \cite{Hao:2021urq}. However the other two actions are somewhat unique as the unfixed vielbein directly appears in them and is absolutely crucial in making them invariant under the transformations in question. In this sense, these are new classes of BMS$_3$ invariant actions, where unlike relativistic theories, fixing background fields doesn't suffice in fixing all ingredients of the geometry.

\subsection{Invariance under tangent space transformations}\label{sec:actions-local-invariance}
We also need our actions to be invariant under tangent space Carroll boosts.
The invariance of $e^{\mu}_0$ under tangent space transformations \eqref{zweibeins-local-transformations}, along with $e^x_0 = 0$ implies that the action $S_{00}$ is automatically invariant under tangent space transformations.

\medskip

However the inverse zweibein $e^{\mu}_1$ is not invariant under local transformations \eqref{zweibeins-local-transformations}. So for the actions $S_{01}$ and $S_{11}$ to be invariant, we require that the derivatives of the scalar field transform as
\begin{equation}\label{tanspace}
	\delta_{\lambda}(\partial_t\Phi) = 0, \quad 	\delta_{\lambda}(\partial_x\Phi) = \lambda_x\partial_t\Phi
\end{equation}
under local transformations, where $\lambda_x = e^1_x\lambda_1$. Naively, at a first glance, it may seem odd that the derivatives of the scalar field, which are spacetime quantities, transform non-trivially under tangent space transformations. Indeed, in the frame formalism, the ordinary derivatives do not transform under local transformations. However we remind the reader that in the composite formalism, the ordinary derivatives do transform under local transformations, as we have seen in \eqref{derivative-local-transf}. This can be interpreted to mean that in the composite formalism, where we have only Carroll diffeomorphisms in the spacetime, the ordinary derivatives are not tensors with respect to Carroll diffeomorphisms. The correctly defined tensors in the Carrollian sense are
\begin{equation}\label{covdev}
	\hat{\partial}_t\Phi = \tau^{\mu}\partial_{\mu}\Phi, \quad \hat{\partial}_x\Phi = (\partial_x - \tau_x\tau^{\mu}\partial_{\mu})\Phi.
\end{equation}
Under Carroll diffeomorphisms $t' = t'(t,x)$, $x' = x'(x)$, we see that $\hat{\partial}_t\Phi$ and $\hat{\partial}_x\Phi$ transform as Carroll scalar and Carroll vector, respectively, with respect to the base space, \ie
\begin{equation}
	(\hat{\partial}_t\Phi)' = \hat{\partial}_t\Phi, \quad (\hat{\partial}_x\Phi)' = \frac{\partial x}{\partial x'}\hat{\partial}_x\Phi.
\end{equation}
Then, as should be the case, we see that the Carroll covariant derivatives above, being true tensors in Carrollian sense in our formalism, are invariant under tangent space transformations provided the ordinary derivatives transform as given in \eqref{tanspace}. Note that for the flat Carroll choice of $\t^\mu= (1,0)$, the covariant derivative $\hat{\partial}_t\Phi$ is simply $\partial_t\Phi$.

\medskip

We can now use the Carroll covariant derivatives defined in \eqref{covdev} to rewrite the three actions $S_{00}$, $S_{01}$ and $S_{11}$ such that these exhibit both Carroll diffeomorphism invariance and local Carroll boost invariance manifestly. This is done by using the completeness relations $\tau_{\mu}\tau^{\mu} = 1$, $\tau_{\mu}e^{\mu} = 0$ along with $\tau^x = 0$ in \eqref{threeactions} to get
\begin{eqnarray}\label{hatted-actions}
	&& S_{00} = \int dt dx\, e\, \hat{\partial}_t\Phi \hat{\partial}_t\Phi, \nonumber \\
	&& S_{01} = \int dt dx\, e\, \hat{\partial}_t\Phi e^x_1\hat{\partial}_x\Phi, \nonumber \\
	&& S_{11} = \int dt dx\, e\, h^{xx}\hat{\partial}_x\Phi\hat{\partial}_x\Phi.
\end{eqnarray}
Here $e^x_1$ is a Carroll vector and $h^{xx}=e^x_1e^x_1$ is a Carroll tensor. Under Carroll diffeomorphisms $t' = t'(t,x)$, $x' = x'(x)$, these transform as
\begin{equation}
	e'^x_1 = \frac{\partial x'}{\partial x}e^x_1, \quad h'^{xx} = \frac{\partial x'}{\partial x}\frac{\partial x'}{\partial x}h^{xx},
\end{equation}
thus making the actions $S_{01}$ and $S_{11}$  in \eqref{hatted-actions} manifestly covariant under Carroll diffeomorphisms. Under local Carroll boosts parametrized by $\lambda_1$, we see from \eqref{zweibeins-local-transformations}, \eqref{2d-h-local-transformations} (using $\tau^x = 0$) that
\begin{equation}
	\delta_{\lambda_1} e^x_1 = 0, \quad \delta_{\lambda_1}h^{xx} = 0,
\end{equation}
thus making the actions $S_{01}$ and $S_{11}$  in \eqref{hatted-actions} manifestly invariant under local Carroll boosts.

\medskip

We would like to mention that, though rewriting in terms Carroll covariant derivatives brings out the symmetries manifestly, we will continue to use the vielbeins and the ordinary derivatives in conjunction in the rest of the paper, for consistency of the notation. While doing so, we will keep in mind the appropriate transformation properties of the vielbeins and the ordinary derivatives under Carroll diffeomorphisms and local Carroll boosts, as given in the preceding discussions.

\medskip

As an aside, this also leads us to the question of tangent space invariance in higher dimensions. Let us consider a naive generalization of our actions to $(d+1)$ dimensions:
\begin{eqnarray}
	S &=& W^{00}\int dtd^dx\, e\, e^{\mu}_0e^{\nu}_0\,\partial_{\mu}\Phi\partial_{\nu}\Phi + 2W^{0a}\int dtd^dx\, e\, e^{\mu}_0e^{\nu}_a\,\partial_{\mu}\Phi\partial_{\nu}\Phi \nonumber \\
	&& \quad + W^{ab}\int dtd^dx\, e\, e^{\mu}_ae^{\nu}_b\,\partial_{\mu}\Phi\partial_{\nu}\Phi \nonumber \\
	&\equiv& W^{00}S_{00} + 2W^{0a}S_{0a} + W^{ab}S_{ab}. \label{scalar-action-higher-dim}
\end{eqnarray}
In $(d+1)$ dimensions, we have both local Carroll boost and local rotations in the tangent space. From \eqref{vielbeins-local-transformations}, we see that $e^{\mu}_0$ is invariant under local Carroll boost and local rotations. Thus we get that the action $S_{00}$ is invariant under tangent space transformations in higher dimensions as well, with $W^{00}$ an arbitrary constant. Requiring local Carroll boost invariance of $S_{ab}$ we get the generalization of the transformations for the derivatives of the scalar field \eqref{tanspace} to
\begin{equation}
	\delta_{\lambda}(\partial_t\Phi) = 0, \quad 	\delta_{\lambda}(\partial_i\Phi) = \lambda_i\partial_t\Phi; \quad \lambda_i = \lambda_a e^a_i. \label{scalar-local-transformation-higher-dim}
\end{equation}
This generalization to higher dimensions can be seen as coming from the invariance under tangent space transformations of general Carroll-covariant derivatives\footnote{Using the identification \eqref{vielbeins-RP-metric}, we note that these Carroll covariant derivatives were also introduced in \cite{Ciambelli:2018ojf} in an alternate formalism in terms of the metric data $(\omega,b_i,a_{ij})$.}
\begin{equation}
	\hat{\partial}_t\Phi = \tau^{\mu}\partial_{\mu}\Phi, \quad \hat{\partial}_i\Phi = (\partial_i - \tau_i\tau^{\mu}\partial_{\mu})\Phi,
\end{equation}
which again transform as Carroll scalar and Carroll vector, respectively, under Carroll diffeomorphisms:
\begin{equation}
	(\hat{\partial}_t\Phi)' = \hat{\partial}_t\Phi, \quad (\hat{\partial}_i\Phi)' = \frac{\partial x^j}{\partial x'^i}\hat{\partial}_j\Phi.
\end{equation}

\medskip

From \eqref{vielbeins-local-transformations} and \eqref{metric-local-transformations}, we see that though the vielbein $e^{\mu}_a$ transforms under local rotations, $h^{\mu\nu}=e^{\mu}_a e^{\nu}_b\delta^{ab}$ is invariant under local rotations. Thus identifying $W^{ab} = \delta^{ab}$, we see that the action
\begin{equation}
	W^{ab} S_{ab} = \int dtd^dx\, e\, h^{\mu\nu}\,\partial_{\mu}\Phi\partial_{\nu}\Phi
\end{equation}
i.e. the higher dimensional analog of the spacelike action, is also invariant under tangent space transformations.

\medskip

Now let us turn to the remaining action $S_{0a}$. Using \eqref{scalar-local-transformation-higher-dim}, we see that $S_{0a}$ is local boost invariant. However $S_{0a}$ is not local rotation invariant due to non-trivial transformation of $e^{\mu}_a$. The only way to have local rotation invariance is to take $W^{0a}$ as a function $W^a(t,x^i)$ such that it transforms as a vector in the tangent space under local rotations but remains invariant under local boosts, \ie\ $\delta W^{a} = \lambda^a_{\ b}W^b$. Then the action in question becomes:
\begin{equation}
	W^{0a}S_{0a} \rightarrow \int dtd^dx\, e\, W^{a} e^{\mu}_0e^{\nu}_a\,\partial_{\mu}\Phi\partial_{\nu}\Phi.
\end{equation}

\smallskip 

\noindent We conclude that we can write $S_{01}$ as a `minimally' coupled action for a scalar field on a Carroll background in $2$-dimensions. However, we cannot write $S_{0a}$ as a meaningful action in higher dimensions without including the vector field $W^a(t,x^i)$ in the action. In this sense, our BMS$_3$ invariant mixed-derivative action in two dimensions is rather intriguing and unique.

\medskip

\subsection{Comments on fixing $e^t_1$}\label{sec:reduced-BMS}

What now remains is to ponder over the unfixed vielbeins in our theory. From a purely geometric perspective, \ie\ for pure Carroll gravity, $e^t_1$ is completely arbitrary for the flat Carroll manifold. For \eg,\  choosing any value for $e^t_1$ doesn't affect the curvatures; those are still vanishing. However when matter is coupled to the flat Carroll spacetime, the complete arbitrariness of $e^t_1$ is lost. We saw that for unfixed $e^t_1$, its non-trivial transformation is essential to ensure BMS invariance of the mixed-derivative and the spacelike actions for the scalar field. However upon fixing $e^t_1$ to a constant value, the resultant mixed-derivative and spacelike actions are not BMS invariant in general. Further, $e^t_1 =0$ turns out to be special compared to other non-zero constant values, where we see certain differences in the symmetries and dynamics of the mixed-derivative and spacelike theories. We will elaborate on this in later sections.

\medskip

For now, let us focus on the reduced symmetries of the scalar field actions after fixing $e^t_1$ to constant. The reduced set of transformations which keep these resultant actions invariant are those that do not change the fixed value of $e^t_1$. From \eqref{et1-transformation}, these reduced transformations should satisfy $\partial_x\xi^t = 0$. Solving this extra condition along with \eqref{cct}, we get that the solution for $\xi$ turns out to be
\begin{equation}
\xi^{\mu} = \lambda x^{\mu} + \kappa^{\mu},
\end{equation}
where $\lambda$ and $\kappa^{\mu}$ are constants describing scaling and translations respectively. Notice that with this choice, the boost spin connection $\Omega_{\mu}^1 = 0$, effectively eliminating all gauge connections on the manifold.

\medskip

To see that the resultant actions have only translations and scaling as the reduced symmetries, let us look at the extra terms generated in the variation of $S_m$ and $S_{sp}$ under BMS transformations \eqref{cct}:
\begin{equation}
	\delta_{\xi}S_m \propto \int dt dx\, \partial_x\xi^t~\dot{\Phi}^2, \quad 	\delta_{\xi}S_{sp} \propto \int dt dx\, \partial_x\xi^t~\dot{\Phi}(\Phi'+ k\dot{\Phi}).
\end{equation}
With no further constraints on $\Phi$, the invariance of the actions $S_m$ and $S_{sp}$ obviously requires $\partial_x\xi^t = 0$.

\medskip

Thus we see that for a constant fixed $e^t_1$, the residual symmetries of $S_m$ and $S_{sp}$ are reduced from BMS$_3$ to only scaling and translations. Interestingly, as we will argue later, the BMS$_3$ invariance for the $S_{sp}$ case can be restored by going to the correct frame defined by the right $\t^\mu$. More intriguingly, following  \cite{Henneaux:2021yzg}, we see that in the mixed action, $e^t_1$ can be thought of as having an additional gauge symmetry. If we take this potentially useful gauge symmetry into consideration, this allows us to fix $e^t_1$ to constant values without actually imposing the extra condition $\partial_x\xi^t=0$ on the Killing field $\xi$. This effective compensation, although unclear from physical perspective but assuming it makes sense physically, may enhance the residual symmetry of the gauge fixed mixed-derivative action. See sec.~\ref{sec:Sm-gauge-symm} for more details on this gauge symmetry.

\subsection{Stress tensors}
The next section onwards we will be turning our attention to the residual symmetries of our three classes of actions with fixed vielbiens, and it is imperative to start with the constraint structure dictated by the corresponding stress tensors. In relativistic $2d$ CFTs, a straightforward consequence of conformal invariance is vanishing of the trace of the stress tensor classically. Analogously in BMS$_3$ invariant theories, the BMS-Weyl symmetry also leads to vanishing of the trace of the stress tensor. As one of our primary goals is to discuss these actions in connection with the tensionless strings, to keep the context clear, from here onwards we choose to work on (Carrollian) cylinder parametrized by $(\tau,\sigma)$ as coordinates. Here $\tau$ is the (null) time coordinate and $\sigma$ is the spatial coordinate with periodicity $\sigma\sim \sigma+2\pi$.
For the combined action \eqref{total-action}, the stress tensor is defined as a variation with respect to the zweibeins 
\begin{equation}
	T^{\mu}_{\ \ \nu} = \frac{e^{\mu}_A}{2 e}\frac{\delta S}{\delta e^{\nu}_A} \label{stress-tensor-definition}
\end{equation}
which explicitly gives
\begin{eqnarray}
	T^{\mu}_{\ \ \nu} &=& W^{AB}\Big(e^{\mu}_A e^{\lambda}_B\partial_{\lambda}\Phi\partial_{\nu}\Phi - \frac{\delta^{\mu}_{\nu}}{2}e^{\rho}_A e^{\lambda}_B\partial_{\rho}\Phi\partial_{\lambda}\Phi \Big). \label{total-stress-tensor}
\end{eqnarray}
On the flat Carroll background \eqref{flat-Carroll-zweibeins} now taken to be the flat Carrollian cylinder, the coefficients of $W^{AB}$ in \eqref{total-stress-tensor} give stress tensors for the three actions in \eqref{gauge-fixed-actions} as follows.

\medskip

\noindent $1$.~Coefficient of $W^{00}$ gives stress tensor for the timelike action\,:
\begin{eqnarray}
	T^\t_{\ \t} = - T^\s_{\ \s} = \frac{1}{2}(\partial_\t\Phi)^2, \quad T^\t_{\ \s} = \partial_\t\Phi\partial_\s\Phi, \quad T^\s_{\ \t} = 0. \label{stress-tensor-timelike}
\end{eqnarray}

\noindent $2$.~Coefficient of $2W^{01}$ gives stress tensor for the mixed derivative action\,:
\begin{eqnarray}
	T^\t_{\ \t} = - T^\s_{\ \s} = \frac{1}{2}e^\t_1(\partial_\t\Phi)^2, \quad T^\t_{\ \s} = \frac{1}{2}(\partial_\s\Phi)^2 + e^\t_1\partial_\s\Phi\partial_\t\Phi, \quad T^\s_{\ \t} = \frac{1}{2}(\partial_\t\Phi)^2. \label{stress-tensor-mixed}
\end{eqnarray}

\noindent $3$.~Coefficient of $W^{11}$ gives stress tensor for the spacelike action\,:
\begin{eqnarray}
	&& T^\t_{\ \t} = - T^\s_{\ \s} = \frac{1}{2}\(-(\partial_\s\Phi)^2 +(e^\t_1)^2(\partial_\t\Phi)^2\), \nonumber \\
	&& T^\t_{\ \s} = e^\t_1\partial_\s\Phi(\partial_\s\Phi + e^\t_1\partial_\t\Phi), \qquad T^\s_{\ \t} = \partial_\t\Phi(\partial_\s\Phi + e^\t_1\partial_\t\Phi). \label{stress-tensor-spacelike}
\end{eqnarray}
Keeping in with the spirit of Carollian structures, we can see the Energy-Momentum tensors are not symmetric in spacetime indices to begin with. But also we would like to remind the reader that for a Carroll boost invariant theory one would require $T^\s_{~\t} = 0$ on shell,\footnote{We would like to note that using the equation of motion for $e^{\tau}_1$, \ie\ $\partial_{\tau}\Phi(\partial_{\sigma}\Phi  + e^{\tau}_1\partial_{\tau}\Phi)=0$, the component $T^{\sigma}_{\ \tau}$ in \eqref{stress-tensor-spacelike} vanishes.} see, \eg\ \cite{deBoer:2021jej} for a nice discussion on this.\footnote{In contrast for a Galilean boost invariant theory one would demand  $T^\t_{~\s} = 0$.} 
In the rest of the paper, we will be using these stress tensors to study the residual symmetries of our three classes of actions. As described before, these symmetries will emerge in the later two cases when we gauge fix $e^\t_1$ to a constant value. 

\bigskip

%%%%%%%%%%%%%%%%%%%%%%%%%%%%%%%%%%%%%%%%%%%%%%
%%%%%%%%%%% SEC: TIMELIKE THEORY %%%%%%%%%%%%%

\section{Warming up: Residual symmetries of the timelike action}\label{sec:timelike}
We will start our discussion with the simplest example, the $2d$ Conformal Carroll model for free scalar field \cite{Hao:2021urq},  which is our timelike action $S_{00}$ in  \eqref{gauge-fixed-actions}. On a flat Carrollian cylinder parametrized by coordinates $(\tau,\sigma)$, the action simply reads
\be{}
 S_t = \int d\t d\s (\partial_{\t}\Phi)^2. \label{St-cylinder}
\ee
\noindent Note that this action can be interpreted as direct Carrollian limit $(c\to 0 ~\text{or}~ \s\to \s, \tau \to \e \tau,~\e\to0)$ on a $2d$ scalar CFT action. The equation of motion in the cylindrical coordinates is:
\begin{equation}
	\partial_{\tau}^2\Phi = 0. \label{timelike-eom}
\end{equation}
Assuming periodic boundary conditions, the general solution in terms of mode expansions can be written as,
\begin{equation}\label{timemode}
	\Phi(\tau,\sigma) = \phi_0 + A_0\sigma + B_0\tau + \sum_{n\neq 0}\frac{i}{n}(A_n - in\tau B_n)e^{-in\sigma},
\end{equation}
where periodicity along $\sigma$ requires $A_0 = 0$, giving the periodic solution
\begin{equation}\label{tlikesol}
	\Phi(\tau,\sigma) = \phi_0 + B_0\tau + \sum_{n\neq 0}\frac{i}{n}(A_n - in\tau B_n)e^{-in\sigma}.
\end{equation}
The mode expansion for the conjugate momentum $\Pi = \partial_{\tau}\Phi$ is
\begin{equation}
	\Pi(\tau,\sigma) = \sum_n B_n e^{-in\sigma}. \label{timelike-momentum}
\end{equation}
It is straightforward to directly check that this action \eqref{St-cylinder} is manifestly  invariant under BMS transformations (\ref{bmstrns}).

\medskip

As mentioned earlier, the timelike action \eqref{St-cylinder} represents a single scalar field version of the gauge-fixed tensionless string action \cite{Bagchi:2013bga,Bagchi:2015nca, Bagchi:2020fpr}. Thus, from a string theory perspective, we can interpret this action as describing a Carroll string on a flat Carroll worldsheet geometry, which is null by construction. This gauge-fixing of the worldsheet geometry leads to the constraints $T^{\mu}_{\ \nu} = 0$. From the expression of the stress tensor \eqref{stress-tensor-timelike}, these constraints can be written as,
\begin{equation}
T^{\mu}_{\ \nu} = 0 \implies T_1\equiv T^{\tau}_{\ \sigma} = 0,~~  T_2\equiv T^{\tau}_{\ \tau} = 0. \label{timelike-constraints}
\end{equation}
Written in terms of oscillator modes from \eqref{timemode}, these constraints become
\begin{eqnarray}
	&& (\partial_{\tau}\Phi)^2 = \sum_{n,m} B_{-m}B_{n+m} e^{-in\sigma} = \sum_n M_n e^{-in\sigma} = 0, \nonumber \\
	&& \partial_{\tau}\Phi\partial_{\sigma}\Phi = \sum_{n,m}(A_{-m} - in\tau B_{-m})B_{n+m} e^{-in\sigma} = \sum_n (L_n - in\tau M_n)e^{-in\sigma} = 0,
\end{eqnarray}
where the modes of the stress tensor are bilinears of the oscillators,
\begin{equation}
	L_n = \frac{1}{2}\sum_m A_{-m}B_{m+n}, \quad M_n = \frac{1}{2}\sum_m B_{-m}B_{m+n}.
\end{equation}

\medskip

\noindent We find the algebra of oscillators by imposing the canonical Poisson's brackets between $\Pi(\tau,\sigma)$ and $\Phi(\tau,\sigma)$. For Carrollian theories, we employ the equal time Poisson's brackets
\begin{eqnarray}
	&& \{\Phi(\tau,\sigma), \Pi(\tau,\sigma')\}_{PB} = \delta(\sigma-\sigma'), \nonumber \\
	&& \{\Phi(\tau,\sigma), \Phi(\tau,\sigma')\}_{PB} = \{\Pi(\tau,\sigma), \Pi(\tau,\sigma')\}_{PB} = 0. \label{canonical-Poisson-brackets}
\end{eqnarray}
It is easy to show they imply the algebra
\begin{equation}\label{timelike-AB-algebra}
	\{A_n,B_m \}_{PB}= -2in\delta_{n+m,0} \quad \{A_n,A_m \}_{PB}= 0, \quad \{B_n,B_m \}_{PB} = 0.
\end{equation}
Bear in mind, this isn't the usual oscillator algebra of scalar field modes, instead they look like those belonging to a particle theory.
Using the canonical Poisson brackets, the algebra of generators $L_n$ and $M_n$ turns out to be
\begin{equation}\label{bms1}
	\{L_n, L_m\}_{PB} = -i(n-m) L_{n+m}, \quad \{L_n, M_m\}_{PB} = -i(n-m) M_{n+m}, \quad \{M_n, M_m\}_{PB} = 0.
\end{equation}
This is the classical part of the BMS$_3$ algebra, which is indeed the residual symmetry algebra for this action. We can identify $L_n$ and $M_n$ as superrotation and supertranslation generators respectively.

\medskip

A lot of discussions have appeared in the literature related to the symmetries of this action, including quantum versions thereof, especially in relation to worldsheet symmetries of null string theories. Here we won't go into the details, stopping only at the classical symmetry analysis. Readers are directed to \cite{Bagchi:2019cay, Bagchi:2020fpr, Bagchi:2020ats, Bagchi:2021rfw, Bagchi:2021ban} for further reading.

\medskip

Complementing the preceeding analysis, where we studied the timelike action \eqref{St-cylinder} from a string theory perspective, we can independently study the same action as a field theory describing a scalar field minimally coupled to a flat Carroll spacetime, \eg\ as discussed in \cite{Hao:2021urq}. Let us briefly comment on this point of view, again taking the background to be a flat Carroll cylinder. Then the equation of motion and the solution are as given in \eqref{timelike-eom}-\eqref{timelike-momentum}. We can vary the action \eqref{St-cylinder} under the infinitesimal BMS transformations \eqref{bmstrns} and use equations of motion to get the conserved Noether currents
\begin{equation}
j^{\tau} = \frac{1}{2}(\partial_{\tau}\Phi)^2\xi^{\tau} + \partial_{\tau}\Phi\partial_{\sigma}\Phi\xi^{\sigma}, \quad j^{\sigma} = (\partial_{\tau}\Phi)^2\xi^{\sigma}.
\end{equation}
We integrate the temporal component $j^{\tau}$ over the $\sigma$-cirlce, and use the mode expansion \eqref{tlikesol} along with the Fourier expansions $f(\sigma) = \sum_n a_n e^{in\sigma}$, $g(\sigma) = \sum_n b_n e^{in\sigma}$ in $\xi^{\tau}$, $\xi^{\sigma}$ \footnote{The killing vectors on a cylinder are $\xi^\t = f'(\s)\t+g(\s),~~\xi^\s = f(\s)$.} to get the conserved charge
\begin{equation}
Q = \int d\sigma j^{\tau} = \sum_n a_n L_n + b_n M_n.
\end{equation}
Then looking at \eqref{bms1}, we can conclude that the above conserved charge generates BMS$_3$ algebra, which are the symmetries of our timelike action \eqref{St-cylinder}. With this vanilla example out of our way, we can then move onto more involved cousins of this action in the next sections.

\bigskip

%%%%%%%%%%%%%%%%%%%%%%%%%%%%%%%%%%%%%%%%%%%%%%
%%%%%%%%%%% SEC: SPACELIKE THEORY %%%%%%%%%%%%

\section{Symmetries of the Spacelike action}\label{sec:spacelike}

In this section, we investigate the action $S_{11}$ in \eqref{threeactions} from a string theory perspective with the geometrical fields being dynamical. As in the usual string theory, we gauge-fix the worldsheet geometry to flat Carroll cylinder $\tau^{\mu}=(1,0)$, $h_{\mu\nu} = diag(0,1)$. This reduces $S_{11}$ to the spacelike action given in \eqref{gauge-fixed-actions}, now in cylindrical coordinates $(\tau,\sigma)$:
\begin{equation}
	S_{sp}=\int d\t d\s(\partial_\s\Phi+e^\t_1\partial_\t \Phi)^2.
\end{equation}

As we saw earlier, this action is BMS$_3$ invariant for arbitrary $e^\t_1(\t,\s)$, which changes in a particular way under BMS transformations. Thus, a priori, we are not allowed to arbitrarily fix $e^{\tau}_1$ to a constant. Nonetheless, we find that for specific constant values of $e^{\tau}_1$, the spacelike action has connections with some known interesting actions in the literature, as we elaborate below. In particular, the value $e^{\tau}_1 = 1$ turns out to be of particular importance, for which the action takes the form
\begin{equation}
	S_{sp}=\int d\t d\s(\partial_\s\Phi+\partial_\t \Phi)^2 = 2\int d\t d\s(\partial_+\Phi)^2, \label{sp-action-et1-fixed}
\end{equation}
where $\sigma^{\pm} = \tau \pm \sigma$. Note the difference with the usual $2d$ CFT action: the holomorphic derivative in the action just drops off here.
This particular class of gauge fixed actions has appeared in many contexts in physics, especially in the study of chiral scalars \cite{Sonnenschein:1988ug} and more recently in Ambitwistor string theories \cite{Casali:2016atr, Casali:2017zkz}, where the action customarily corresponds to the choice $e^\t_1=-1$ instead.

\medskip
The components of the stress tensor with this choice take the form
\begin{equation}
	T^\t_{\ \t} = - T^\s_{\ \s} =  \frac{1}{2}(\dot{\Phi}+\Phi')(\dot{\Phi}-\Phi'), \quad T^\t_{\ \s} = (\dot{\Phi}+\Phi')\Phi', \quad T^\s_{\ \t} = (\dot{\Phi}+\Phi')\dot{\Phi}, \label{sp-et1fixed-stress-tensor-2nd}
\end{equation}
which can be obtained by putting $e^\t_1=1$ in \eqref{stress-tensor-spacelike}. Now let us write the stress tensor in terms of the  momentum conjugate to $\Phi$: 
\begin{equation}
	P = \frac{\delta \mathcal{L}}{\delta\dot{\Phi}} = 2(\dot{\Phi}+\Phi') = \partial_{+}\Phi. \label{spacelike-momentum}
\end{equation}
Then the components of stress tensor in terms of $P$ are
\begin{equation}
	T^\t_{\ \t} = - T^\s_{\ \s} = \frac{P^2}{8} - \frac{P\Phi'}{2}, \quad T^\t_{\ \s} = \frac{P\Phi'}{2}, \quad T^\s_{\ \t} = \frac{P^2}{4} - \frac{P\Phi'}{2}. \label{sp-et1fixed-stress-tensor-1st}
\end{equation}
This stress tensor is traceless and conserved but clearly $T^\s_{\ \t} \neq 0$, which is a telltale sign of Carroll boost invariance being broken. In keeping with the string theory description, we now have to impose the constraints
\begin{equation}
	T^{\mu}_{\ \nu} = 0 \quad \implies \quad P\Phi' = 0, \quad P^2 = 0.
\end{equation}

\subsection{Symmetries in ``lightcone" coordinates}
Although the gauge-fixed spacelike action does not have full Carroll symmetries anymore, we can go to lightcone coordinates to restore Carroll boost invariance in one of the lightcone directions. This is also evident in the chiral structure of the spacelike Lagrangian. Taking a cue from Ambitwistor string literature  \cite{Casali:2016atr, Casali:2017zkz} we write the action in lightcone coordinates on a Carrollian cylinder using,
\begin{eqnarray}
	\sigma^{\pm} = \tau \pm \sigma, \quad \partial_{\pm} = \frac{1}{2}(\partial_{\tau} \pm \partial_{\sigma}). \label{lc-coordinates}
\end{eqnarray}
Now the gauge-fixed spacelike action becomes
\begin{equation}
	S = 2\int d\sigma^+ d\sigma^- (\partial_+\Phi)^2. \label{sp-action-lightcone}
\end{equation}
We see that this action is manifeslty invariant under the BMS$_3$ transformations in lightcone coordinates:
\begin{equation}
	\sigma^{\pm} \rightarrow \sigma^{\pm} + \xi^{\pm}; \qquad \xi^+ = \sigma^+ \partial_- f(\sigma^-) + g(\sigma^-), \quad \xi^- = f(\sigma^-). \label{BMS-transf-lc}
\end{equation}

However note that this isn't equivalent to having BMS$_3$ in $(\t,\s)$ coordinates as the Killing equations are drastically different,
\begin{eqnarray}
	&& \partial_+\xi^+ = \partial_-\xi^- \quad \implies\quad \partial_{\tau}\xi^{\sigma} + \partial_{\sigma}\xi^{\tau} = 0, \nonumber \\
	&& \partial_+\xi^- = 0 \quad \implies\quad \partial_{\tau}\xi^{\tau} - \partial_{\sigma}\xi^{\sigma} - \partial_{\tau}\xi^{\sigma} + \partial_{\sigma}\xi^{\tau} = 0. \label{lc-BMS-in-tx-1}
\end{eqnarray}
We can write the conformal Killing equations as
$
	\partial_{\tau}\xi^{\sigma} = -\partial_{\sigma}\xi^{\tau}, \quad \partial_{\tau}\xi^{\tau} - \partial_{\sigma}\xi^{\sigma} = 2\partial_{\tau}\xi^{\sigma}, \label{lc-BMS-in-tx-2}
$
which are not equivalent to BMS transformations in the original coordinates \eqref{cct}. This happens because the frame defined by (\ref{lc-coordinates}) is {Carroll inequivalent} to the usual $(\t,\s)$  frame we have been working on till now. 

\medskip

An alternative way to look at this action is to consider it to be the ``timelike'' action corresponding to the separate Carroll frame (i.e. choice of $h_{\mu\nu}$ and $\tau^\mu$) we mentioned in \eqref{flat-Carroll-metric-v2}. In this case the new ``timelike'' action in cylindrical coordinates is: 
\be{}
\tilde{S}_{t}\sim \int d\t d\s\, e\, \t^{\mu}\t^{\nu}\,\partial_{\mu}\Phi\partial_{\nu}\Phi = \int d\t d\s(\partial_+\Phi)^2,
\ee
since $\t^\mu = (1,1)$ is the new null vector in this case. This fits in perfectly with what we have already found, i.e. the new action is invariant under BMS$_3$ transformations, albeit in the $\s^\pm$ direction. The (BMS) conformal Killing equations
$\mathcal{L}_{\xi}h_{\mu\nu} = -2\lambda h_{\mu\nu}, ~ \mathcal{L}_{\xi}\tau^{\mu} = \lambda \tau^{\mu}$ explicitly take the form  \eqref{lc-BMS-in-tx-1} for this case.\footnote{This choice of the null vector also relates to a cousin of the ``Ambitwistor'' gauge of null string theories, details of which will be discussed later. } Also notice that for the spacelike action with any constant $e^{\tau}_1 = k,~k>0$ can be re-interpreted as the timelike action in a Carroll frame with $\t^\mu = (1,\frac{1}{k})$, and hence can be shown to be BMS$_3$ invariant in a judiciously chosen frame where the Carroll boost invariance is automatically restored \footnote{ For more details on this choice of frame one can refer to the appendix \ref{ApA}.}. Note here, in these cases the action could be rewritten, upto a constant, using the timelike covariant derivative \eqref{covdev},
\be{}
\tilde{S}_{t}\sim\int d\t d\s(\hat{\partial}_{\t}\Phi)^2,
\ee
which guarantees invariance under BMS$_3$ transformations since $\hat{\partial}_t\Phi$ always transforms as a Carroll scalar. 
However with the choice of $e^{\tau}_1 = 0$, this argument does not work as the null vector then has divergent components. In fact one can show with $e^{\tau}_1 = 0$ we get a purely Galilean covariant action. This is also supported by the fact that with this choice, the stress tensor component $T^\t_{~\s}$ clearly vanishes \eqref{stress-tensor-spacelike}, implying an underlying Galilean takeover. However in two dimensions, Galilean and Carrollian conformal algebras are classically isomorphic to each other, so even at this singular point, the form of the symmetry algebra remains unchanged.

\medskip

Coming back to the lightcone case, using the transformation $\Sigma^{\mu}_{\ \nu}(\sigma^+,\sigma^-) = \frac{\partial\sigma^{\mu}}{\partial x^{\alpha}}\frac{\partial x^{\beta}}{\partial\sigma^{\nu}}T^{\alpha}_{\ \beta}$, the components of the stress tensor \eqref{sp-et1fixed-stress-tensor-2nd} can be written in lightcone coordinates as
\begin{equation}
	\Sigma^{+}_{\ +} = - \Sigma^{-}_{\ -} = 2(\partial_+\Phi)^2, \quad \Sigma^{+}_{\ -} = 4\partial_+\Phi\partial_-\Phi, \quad \Sigma^{-}_{\ +} = 0. \label{sp-EM-lc}
\end{equation}
One can clearly see that $\Sigma^{-}_{\ +}$ vanishes here, signalling once again the restoration of Carroll boost invariance in these coordinates.
In lightcone coordinates, the conjugate momentum is $P = 4\partial_+\Phi$, which dictates the form of stress tensors as follows:
\begin{equation}
	\Sigma^{+}_{\ +} = - \Sigma^{-}_{\ -} = \frac{P^2}{8}, \quad \Sigma^{+}_{\ -} = \frac{P^2}{4} - P\Phi', \quad \Sigma^{-}_{\ +} = 0. \label{sp-EM-lc-P}
\end{equation}
From \eqref{sp-EM-lc-P}, we see that imposing $\Sigma^{\mu}_{\ \nu} = 0$ gives usual ``null string" constraints
\begin{equation}
	P\Phi' = 0, \quad P^2 = 0.
\end{equation}
However from \eqref{sp-EM-lc}, we get the constraints
\begin{equation}
	\partial_+\Phi\partial_-\Phi = 0, \quad (\partial_+\Phi)^2 = 0, \quad \ie\quad P\partial_-\Phi = 0, \quad P^2 = 0,
\end{equation}
respectively. The constraint $P\Phi' = 0$ here appears to be just a combination of $P\partial_-\Phi = 0$ and $P^2 = 0$, hence isn't an independent object.  
\subsection{Mode expansion and charges}\label{periodi}
In lightcone coordinates, the Euler-Lagrange equation of motion coming from (\ref{sp-action-lightcone}) reads
\begin{equation}
	\partial^2_+\Phi=0,
\end{equation}
and the Hamilton's equations are $P = 4\partial_+\Phi$, $\partial_+ P = 0$. The general solution for $\Phi$ associated to these equations is, 
\begin{equation}
	\Phi(\sigma^+,\sigma^-) = A(\sigma^-) + \frac{\sigma^+}{4} P(\sigma^-),
\end{equation}
where the arbitrary functions have mode expansions
\begin{equation}
	A(\sigma^-) = \sum_n A_n e^{-i n \sigma^-}, \quad P(\sigma^-) = \sum_n P_n e^{-i n \sigma^-}.
\end{equation}
Note that this solution is not periodic in $\sigma$ (or in $\sigma^{\pm}\rightarrow \sigma^{\pm}\pm 2\pi$). A similar problem for the modes was noticed in \cite{Casali:2016atr}, and was remedied by introducing the weak periodicity conditions
\be{}
P(\s^- + 2\pi) = P(\s^-),~~A(\s^- + 2\pi) \approx A(\s^-)-\frac{\pi}{2} P(\s^-),
\ee 
which imposes a condition on the $A_n$ modes as
\be{}
A_n \approx A_n-\frac{\pi}{2}  P_n.
\ee
This identification clearly does not hamper the canonical Poisson brackets imposed on the system that reads,
\begin{equation}
	\{A_n,P_m\}=\delta_{n+m,0}, \quad \{A_n,A_m\}=0, \quad \{P_n,P_m\}=0.
\end{equation}
But this is a subtle problem related to the actual Ambitwistor gauge that cannot be addressed in the second order equations of motion, so the confusion regarding the interpretation of the non-periodic solution persists. We will however take these equations at the face value and proceed for now, and a detailed discussion will be given in section \ref{sec:ilst}. 
The constraints coming from the lightcone stress tensor are
\begin{eqnarray}
	\Sigma^+_{\ +} \sim P^2 &=& \sum_{n,m} P_{-m}P_{n+m}e^{-in\sigma^-} \sim \sum_n M_n e^{-in\sigma^-}, \\
	\Sigma^+_{\ -} = P\partial_-\Phi &=& \sum_{n,m}\Big( imA_{-m}P_{n+m} - \frac{in\sigma^+}{8}P_{-m}P_{n+m} \Big)e^{-in\sigma^-} \nonumber \\
	&=& \sum_n(L_n - in\sigma^+M_n)e^{-in\sigma^-}. 
\end{eqnarray}
We see that even with the non-periodic solution, we can write the constraint $P\partial_-\Phi$ in the desired form of BMS$_3$ stress tensor. The above constraints again generate the classical part of the BMS$_3$ algebra \eqref{bms1}. We also notice the modes of the second constraint $(L_n - in\sigma^+M_n)$ generate conformal transformations along the $\s^-$, bringing our argument throughout this section to a full circle.

%%%%%%%%%%%%%%%%%%%%%%%%%%%%%%%%%%%%%%%%%%%%%%
%%%%%% SEC: MIXED-DERIVATIVE THEORY %%%%%%%%%%

\newpage

\section{The Mixed-derivative theory}\label{sec:mixed}

In this section, we concentrate on one of the main findings of this work, the covariant mixed-derivative action $S_m$ in \eqref{gauge-fixed-actions}. This one will be technically different from the other two cases we have discussed in the preceding sections. We will see the fate of the symmetries when we fix the unspecified vielbein  $e^\t_1$ in this case. As discussed before, fixing $e^\t_1$ reduces the conformal Carroll symmetries down to scaling and translations only. However as we also mentioned earlier, a gauge symmetry of $e^\t_1$ can potentially compensate this effect and bring back boost invariance to the system. 

\subsection{Gauge symmetry of $e^\t_1$}\label{sec:Sm-gauge-symm}
In sec.~\ref{mainsec}, we saw that the mixed derivative action, given by 
\begin{equation}\label{Sm}
	\ S_{m} = \int d\t d\s \Big( \partial_\s\Phi \partial_\t\Phi + e^\t_1(\partial_\t\Phi)^2 \Big)
\end{equation}
with arbitrary $e^\t_1(\t,\s)$ is  manifestly BMS$_3$ invariant. Now to see the aforementioned gauge symmetries associated to the unfixed zweibein $e^\t_1$, we need to recast $S_{m}$ into first order form. The conjugate momentum to $\Phi$ in this case is,
\begin{equation}\label{conjuate-momentum}
	\Pi = \frac{\delta\mathcal{L}_{m}}{\delta \dot{\Phi}} = 2e^\t_1\dot{\Phi} + \Phi'
\end{equation}
where $\dot{\Phi} = \partial_\t\Phi$ and $\Phi' = \partial_\s\Phi$. The Hamiltonian is:
\begin{equation}
	\mathcal{H}_{m} = \Pi\dot{\Phi} - \mathcal{L}_{m} = e^\t_1\dot{\Phi}^2 = \frac{(\Pi - \Phi')^2}{4e^\t_1}.
\end{equation}
 We can write the first order action using the  Hamiltonian  as
\begin{equation}
	S^{H}_{m} = \int d^2\s \Big( \Pi\dot{\Phi} -  \frac{(\Pi - \Phi')^2}{4e^\t_1} \Big).
\end{equation}
Since $e^\t_1$ is arbitrary in this action, we redefine it as $\theta = \frac{1}{e^\t_1}$. Then the Hamiltonian action takes the form, 
\begin{equation}
	S^{H}_{m} = \int d^2\s \Big( \Pi\dot{\Phi} -  \frac{\theta}{4}(\Pi - \Phi')^2 \Big).
\end{equation}
Now we remind ourselves, generically  the Hamiltonian action for gauge theories is written as
\begin{equation}
	I_H = \int d^2\s \big(\Pi\dot{\Phi} - H_0 + \lambda^a\phi_a \big),
\end{equation}
where $H_0$ is the canonical Hamiltonian and  $\lambda^a$  are Lagrange multipliers that imposes the constraints $\phi ^a=0$. For our case, $\theta$ is the only  Lagrange multiplier and the corresponding constraint is given by $\phi =\frac{1}{4} (\Pi - \Phi')^2=0$. The canonical Hamiltonian $H_0$ is also equal to zero in this case.  Using the Poisson brackets
\begin{eqnarray}
[\Phi(\t,\s), \Pi(\t,\s')] = \delta(\s-\s'), \quad [\Phi(\t,\s), \Phi(\t,\s')] = [\Pi(\t,\s), \Pi(\t,\s')] = 0,
\end{eqnarray}
it can be verified that the constraint is indeed a first class constraint and hence will generate gauge symmetries that keep the action invariant. 
For an arbitrary gauge function  $\epsilon(\t,\s)$, the field   $\Phi$ transforms as\footnote{Using the expression \eqref{conjuate-momentum} for the conjugate momentum, we can write the gauge transformation \eqref{scalar-gauge-transformation} of the scalar field as $\delta_{\epsilon}\Phi = 2\epsilon e^{\tau}_1 \partial_{\tau}\Phi$.}
\begin{equation}\label{scalar-gauge-transformation}
	\delta_{\epsilon}\Phi = \int d^2\s' \epsilon(\t,\s')\big[\Phi(\t,\s),\phi(\t,\s')\big] = \epsilon(\t,\s)(\Pi - \Phi')
\end{equation}
and the canonical conjugate $\Pi$ will transform under the gauge symmetry as,
\begin{equation}
	\delta_{\epsilon}\Pi = \int d^2\s' \epsilon(\t,\s')\big[\Pi(\t,\s),\phi(\t,\s')\big] = -2\big(\epsilon(\t,\s)(\Pi - \Phi')\big)'.
\end{equation}
 Then we have the gauge transformations of various terms in the action $S_m^H$ as
\begin{eqnarray}
	&& \delta_{\epsilon}(\Pi\dot{\Phi}) = -\epsilon\dot{\phi} + B = \dot{\epsilon}\phi + B_1, \nonumber \\
	&& \delta_{\epsilon}(\Pi - \Phi')^2 = 4\big((\epsilon\phi)' + \epsilon'\phi \big) \ \implies\ -\frac{1}{4}\theta\delta_{\epsilon}\phi = -\theta\epsilon'\phi - \epsilon\phi\theta' + B_2,
\end{eqnarray}
where $B$, $B_1$, $B_2$ are total derivative terms. Using these expressions, we have the transformation of the total Lagrangian as the following, 
\begin{eqnarray}
	\delta_{\epsilon}\mathcal{L}^H_{m} &=& \delta_{\epsilon}(\Pi\dot{\Phi}) - \frac{1}{4}\theta\delta_{\epsilon}\phi - \frac{1}{4}\delta_{\epsilon}\theta \phi \nonumber \\
	&=& (\dot{\epsilon}-\theta\epsilon' - \epsilon\theta')\phi - \frac{1}{4}\delta_{\epsilon}\theta \phi + \text{boundary terms}.
\end{eqnarray}
In order for  the action to be  gauge invariant, the Lagrange multiplier $\theta$ has to transform as
\begin{equation}
	\delta_{\epsilon}\theta = 4(\dot{\epsilon} - \theta\epsilon' - \epsilon\theta').
\end{equation}
Then using $e^\t_1 = \frac{1}{\theta}$, the gauge transformation of $e^\t_1$ that keeps the action $S_m$ in \eqref{Sm} gauge-invariant is
\begin{equation}\label{et1-gauge-transformation}
	\delta_{\epsilon} e^\t_1 = -(e^\t_1)^2\dot{\epsilon} + e^\t_1\epsilon' - \epsilon(e^\t_1)'.
\end{equation}

\medskip

\noindent \underline{\textbf{Towards a way to gauge fix $e^\t_1 = 1$:}}

\medskip

From \eqref{et1-transformation}, we see that under BMS$_3$ transformations \eqref{cct}, $e^\t_1$ changes as $$\delta_{\xi}e^\t_1 = -\xi^{\nu}\partial_{\nu}e^\t_1 + \partial_\s\xi^\t.$$ That was the reason the action lost invariance under the full BMS$_3$ group when $e^{\t}_1$ assumes a fixed value. Now, following a similar logic presented in \cite{Henneaux:2021yzg}, we could possibly use the gauge symmetries mentioned above and restore BMS invariance by simultaneously making a suitable gauge transformation that will preserve the  condition $e^\t_1 = 1$, i.e,
\begin{equation}
	(\delta_{\xi} + \delta_{\epsilon})e^\t_1 = 0 \ \implies\ (\xi^\t)'+\epsilon'-\dot{\epsilon} = 0.
\end{equation}
From conformal  Killing equations on a flat Carroll background \eqref{cct}-\eqref{bmstrns}, we have $\xi^\t= \t f'(\s) + g(\s)$. Hence the above equation can be solved generically if we choose 
\begin{equation}
\epsilon (\t,\s)=-\big( f'(\s)\t+ g(\s)+f(\s)+k \big).
\end{equation}
Here $f$, $g$ are arbitrary functions of $\s$ and $k$ is a constant, signifying we can always choose an $\e$ of this form. Then we can potentially argue that the gauge-fixed action
\begin{equation}\label{Sm-gaugefixed}
	S_m = \int d\t d\s (\dot{\Phi}^2 + \dot{\Phi}\Phi')
\end{equation}
would be  invariant under the BMS transformation, if we simultaneously make a compensating gauge transformation, \ie\ $(\delta_{\xi} + \delta_{\epsilon})S_m = 0$\footnote{The idea of compensating a diffeomorphism with a gauge transformation works mathematically to give rise to a symmetry enhancement, but the physical implication of such a composition of two physically distinct transformations is not entirely clear to us.}. Remember, appearance of this gauge symmetry does not depend on the constant value of $e^{\tau}_1$ being chosen, and the resulting physics should always be independent of this value.

\medskip

With the offshell symmetries of the mixed-derivative action taken care of, we now put the theory onshell, in what follows. While doing so, we fix $e^{\tau}_1=1$ without using the (compensating) gauge symmetry discussed above, which leads to breaking of symmetries to only translations and scaling, as mentioned before. However we will see that, onshell, there is an enhancement in symmetries to one copy of the Virasoro algebra. This is an interesting feature since it shows similarities with the Floreanini-Jackiw (FJ) chiral scalar field theory \cite{Floreanini:1987as}, and thus leads us to interpret the action \eqref{Sm-gaugefixed} as a time-space inverted version of the FJ action. However we would also like to note that this equivalence between the mixed-derivative and the FJ theories only occurs onshell. Offshell, the FJ action is a relativistic theory unlike the mixed-derivative action, which by construction is a Carrollian theory. Close cousins of this chiral action have appeared in myriad of physical systems, most recently in the arena of near horizon dynamics of scalar fields \cite{Grumiller:2019tyl}.

\subsection{Onshell Symmetries}

In the previous discussions we have seen that upon fixing $e^\t_1$ to a constant value, without using the compensating gauge transformation introduced in sec.~\ref{sec:Sm-gauge-symm}, the action loses the boost invariance and consequently the infinite BMS extensions. However in this section, we show that these symmetries can be partially lifted to one copy of Virasoro when we go onshell.

\medskip

The Euler Lagrange equation of motion from the variation of the gauge-fixed action \eqref{Sm-gaugefixed} is 
 \begin{equation}
    \partial^2_\t\Phi+\partial_\t\partial_\s\Phi=0.
 \end{equation}
As the covariant action has Carrollian diffeomorphism and  Weyl invariance, we are interested in residual conformal Carrollian transformations. The conformal Killing equations on flat Carroll background are given by \eqref{cct}. Since the Weyl weight for the scalars in $(1+1)$ dimensions is zero, the transformations of  $\Phi$ under these conformal Carroll transformations would be 
\begin{equation}
    \delta_{\xi}\Phi=-\xi^{\rho}\partial_{\rho} \Phi
\end{equation}
Now under these transformations satisfying the conformal Killing equations, different terms in the Lagrangian will transform as follows. The $\dot{\Phi}$ term transforms as
\begin{eqnarray}
	\delta_{\xi}[(\partial_\t \Phi)^2]&=& -2 \partial_\t \Phi\delta_{\xi}(\partial_\t \Phi) \\ \nonumber
 	&=& -2 \partial_\t \Phi [(\partial_{\t}\xi^{\t})\partial_{\t}\Phi+\xi^{\t}\partial_{\t}^2\Phi+\xi^{\s}\partial_{\t}\partial_{\s}\Phi] \\ \nonumber
 	&=& -\partial_{\t}[\xi^{\t}(\partial_{\t}\Phi)^2]-\partial_{\s}[\xi^{\s}(\partial_{\t}\Phi)^2]
\end{eqnarray}
and the cross derivative term transforms as
\begin{eqnarray}
	\delta_{\xi}(\partial_\t \Phi\partial_\s\Phi)= -\partial_{\t}[\xi^{\t}(\partial_\t \Phi\partial_\s\Phi)]-\partial_{\s}[\xi^{\s}(\partial_\t \Phi\partial_\s\Phi)]-\partial_{\s}\xi^{\t}(\partial_{\t}\Phi)^2.
\end{eqnarray}
The last term in the above equation seems to be problematic, but it also turns out to be a boundary term if the equations of motion for $\Phi$ is used:
\begin{equation}
    \partial_{\s}\xi^{\t}(\partial_{\t}\Phi)^2= (\partial_\t+\partial_\s)[(\xi^{\t}-\xi^{\s})(\partial_\t \Phi)^2]
\end{equation}
Hence we have the total variation of the Lagrangian
\begin{equation}
    \delta_{\xi} \mathcal{L}= - \partial_{\rho} \Lambda ^{\rho},
\end{equation}
where we have
\begin{eqnarray}
    \Lambda^{\t}=\xi^{\t}[2(\partial_{\t}\Phi)^2+\partial_{\t}\Phi\partial_{\s}\Phi]- \xi^{\s}(\partial_\t\Phi)^2, \quad \Lambda^{\s}=[\xi^{\t}(\partial_\t \Phi)^2+\xi^{\s}\partial_{\t}\Phi\partial_{\s}\Phi].
\end{eqnarray}
Using this expression for $\delta_{\xi}\mathcal{L}$, we compute the onshell expression for the Noether current:
\begin{equation}
    J^{\alpha}=\frac{\partial \mathcal{L}}{\partial(\partial_{\alpha}\Phi)}\delta_{\xi}\Phi-\Lambda^{\alpha},
\end{equation}
whose components are
\begin{equation}
    J^{\t}=\xi^{\s}(\partial_{\t}\Phi+\partial_{\s}\Phi)^2, \quad J^{\s}=0.
\end{equation}
This current is conserved onshell and the charges associated with it can be computed by integrating $J^\t$ over the spatial slice:
\begin{equation} \label{charge}
Q_\xi=\int d\s [\xi^{\s}(\partial_{\t}\Phi+\partial_{\s}\Phi)^2].
\end{equation}
One noteworthy point here is that the dependence on $\xi^\t$ is entirely cancelled out in the expression of charges. This indicates that all the supertranslation charges ($M_n$) vanish onshell, leaving only the superrotation charges ($L_n$) non-trivial. These superrotation charges satisfy one chiral copy of Virasoro algebra. This means the symmetries of the theory are enhanced to chiral Virasoro onshell .
 
\medskip

\noindent Same conclusion can be achieved by improving the stress tensor of the theory as well. The conserved currents associated with  BMS transformations can be constructed using stress tensor components as
\begin{equation}
    J^{\alpha}=T^{\alpha}_{\ \beta} \xi^{\beta}.
\end{equation}
Upon using the conservation of stress tensor and the tracelessness condition, we get from the above equation
\begin{equation}
    \partial_{\alpha}J^{\alpha}=(\partial_{\s}\xi^{\t})T^\s_{\ \t}.
\end{equation}
We see that the conservation of Noether current requires us to have a stress tensor with $T^\s_{\ \t}=0$. In quantum theory this condition on the stress tensor can be accounted for Carrollian boost Ward identity. For our case, the components of the stress tensor \eqref{stress-tensor-mixed} for $e^\t_1=1$ are
\begin{eqnarray}\label{mixed-stress-tensor}
	T^\t_{\ \t} = - T^\s_{\ \s} = \frac{1}{2}(\partial_\t\Phi)^2, \quad T^\t_{\ \s} = \frac{1}{2}(\partial_\s\Phi)^2 + \partial_\s\Phi\partial_\t\Phi, \quad T^\s_{\ \t} = \frac{1}{2}(\partial_\t\Phi)^2.
\end{eqnarray}
This stress tensor is traceless and conserved but $T^\s_{\ \t} \neq 0$. This again implies boost non-invariance but it is possible to  improve the stress tensor and set $T^\s_{\ \t}=0$ by using equations of motion. 

\medskip

In order to do this improvement we add $B^{\alpha}_{\ \beta}$ to $T^{\alpha}_{\ \beta}$ such that $B^\t_{\ \t}=-B^\s_{\ \s}\equiv f_1(\t,\s)$, $B^\t_{\ \s}\equiv f_2(\t,\s)$ and $B^\s_{\ \t}=-(\partial_\t \Phi)^2$. The conservation equations $\partial_{\alpha}B^{\alpha}_{\ \beta} = 0$, since $T^{\alpha}_{\ \beta}$ is conserved, become
\begin{equation}
	\partial_{\t}f_1=\partial_{\t}\Phi\partial_{\t}\partial_{\s}\Phi, \quad \partial_{\t} f_{2}
	= \partial_{\s}f_1.
\end{equation}
A general (not using double derivatives acting on $\Phi$) solution of the first equation above is
 \begin{equation}
 f_1=\frac{1}{2}[\lambda(\partial_\t \Phi+ \partial_\s \Phi)^2-(\partial_\t \Phi)^2],
 \end{equation}
where the $\lambda$-term, $\lambda$ being an arbitrary constant parameter, corresponds to the equation of motion in the conservation equations. With this solution for $f_1$, the second conservation equation becomes
\begin{equation}
   \partial_{\t} f_{2} = \lambda (\partial_{\t}\Phi+\partial_{\s}\Phi)(\partial_{\t}\partial_{\s}\Phi+\partial^2_{\s}\Phi)-\partial_{\t}\Phi\partial_{\t}\partial_{\s}\Phi.
\end{equation}
For simplicity and without loss of generality, we choose $\lambda=0$ and get
\begin{equation}
    f_2=-\frac{1}{2}(\partial_{\t}\Phi)^2.
\end{equation}
This solution for $(f_1,f_2)$ gives us the improved stress tensor $\Sigma^{\mu}_{\ \nu} = T^{\mu}_{\ \nu} + B^{\mu}_{\ \nu}$, whose components are
\begin{equation}\label{improved-stress-tensor-mixed}
	T_1 \equiv \Sigma^\t_{\ \s}=\frac{1}{2}(\partial_{\t}\Phi+\partial_{\s}\Phi)^2, \quad T_2\equiv \Sigma^\t_{\ \t}=-\Sigma^\s_{\ \s}=0, \quad \Sigma^\s_{\ \t} = 0.
\end{equation}
This meets the criteria for constructing conserved currents and matches with what we have seen above from Noether's procedure.

\medskip

We also note that there is another way to obtain the improved stress tensor directly from the variation of the action by making use of the degenerate nature of the spacetime. In particular, owing to the constraint $e^{\mu}_1\tau_{\mu} = 0$, we project out the pure timelike component while taking variation with respect to $e^{\mu}_1$, \ie
\begin{equation}\label{vielbein-modified-variation}
\frac{\delta}{\delta e^{\nu}_1}e^{\mu}_1\partial_{\mu}\Phi = \partial_{\nu}\Phi - \tau_{\nu}\tau^{\mu}\partial_{\mu}\Phi.
\end{equation}
Using this modified variation with respect to $e^{\mu}_1$ in the formula \eqref{stress-tensor-definition} for the stress tensor and putting the flat values of vielbeins \eqref{flat-Carroll-zweibeins} along with $e^\t_1 = 1$ in the resulting expressions, we directly get \eqref{improved-stress-tensor-mixed}. Naively, though, it may seem that this method is different than improving the stress tensor. However, a closer look reveals that the two procedures are indeed equivalent. To see this, we note that while improving the stress tensor to make $T^\s_{\ \t}$ vanishing, we are effectively subtracting out $(\partial_\t\Phi)^2$ (recall that $B^\s_{\ \t} = -(\partial_\t\Phi)^2$). In the second procedure using the modified variation \eqref{vielbein-modified-variation}, projecting out the timelike component effectively amounts to imposing the equation of motion for $e^\t_1$, \ie\ $(\partial_\t\Phi)^2 = 0$. Then putting $(\partial_\t\Phi)^2 = 0$ in \eqref{mixed-stress-tensor} gives the improved stress tensor \eqref{improved-stress-tensor-mixed}. Thus the two procedures are equivalent and give the desired improved stress tensor.

\subsection{Mode expansions and symmetry algebra}

We start with the equation of motion, 
\begin{equation} \label{mixed-eom}
\partial^2_{\tau} \Phi + \partial_{\tau}\partial_\sigma \Phi =0.
\end{equation}
The most general solution ansatz subject to periodic boundary conditions would be 
\begin{equation}
\Phi (\tau,\sigma)= \sum_{n} \Phi_{n}(\tau)e^{-in\sigma},
\end{equation}
where the modes $\Phi_n(\tau)$ satisfy 
\begin{equation}
\partial^2_\tau{\Phi_n(\tau)}-in\partial_\tau{\Phi_n(\tau)}=0.
\end{equation}
These equations are generically solved by 
\begin{equation}
	\Phi_0(\tau)=\phi_0 + B_0\tau, \quad \Phi_n(\tau)=\frac{i}{n}(A_n - B_n e^{in\tau}) \ \forall\ n \neq 0,
\end{equation}
which gives the mode expansion for $\Phi$ as
\begin{equation}\label{mixed-periodic-solution}
	\Phi(\tau,\sigma)=\phi_0 + B_0\tau + i\sum_{n \neq 0}\frac{1}{n}(A_n - B_ne^{in\tau})e^{-in\sigma}.
\end{equation}
Note that the general solution to \eqref{mixed-eom} can be written as,
 $$\Phi(\tau,\sigma)=\phi_0 + \frac{(A_0 + B_0)}{2}\tau + \frac{(A_0 - B_0)}{2}\sigma + i\sum_{n \neq 0}\frac{1}{n}(A_n - B_ne^{in\tau})e^{-in\sigma}.$$ 
 Imposing periodicity in $\sigma$ into the above implies $A_0 = B_0$, giving the periodic solution \eqref{mixed-periodic-solution}.

\medskip

\subsubsection*{\underline{Poisson's brackets and charge algebra:}}

\medskip

\noindent The canonical momenta conjugate to $\Phi$ is 
\begin{equation}
\Pi = \frac{\delta \mathcal{L}}{\delta \dot{\Phi•}}= 2\dot{\Phi}+\Phi',
\end{equation}
where $\dot{\Phi}$ and $\Phi '$ denote $\tau$ and $\sigma$ derivatives of $\Phi$ respectively. In terms of the oscillators
\begin{eqnarray}
	&& \dot{\Phi}(\tau,\sigma)=B_0 + \sum_{n \neq 0} B_n  e^{in(\tau-\sigma)} = \sum_{n} B_n  e^{in(\tau-\sigma)}, \nonumber \\
	&& \Phi'(\tau,\sigma)= \sum_{n \neq 0}(A_n - B_n e^{in\tau})e^{-in\sigma} = \sum_{n}(A_n - B_n e^{in\tau})e^{-in\sigma},
\end{eqnarray}
where we have used $B_0 = A_0$ in the second equality in $\Phi'$ above. Then the expression for the conjugate field momentum becomes
\begin{equation}
	\Pi(\tau,\sigma)= \sum_{n} (A_n + B_n e^{in\tau})e^{-in\sigma}.
\end{equation}
We find the algebra of oscillators by imposing the canonical Poisson's brackets between $\Pi(\tau,\sigma)$ and $\Phi(\tau,\sigma)$. For Carrollian theories, the equal time Poisson's brackets \eqref{canonical-Poisson-brackets} imply the algebra for oscillators:
\begin{equation}\label{mixed-AB-algebra}
	\{A_n,A_m \}_{PB}= -in\delta_{n+m,0} \quad \{B_n,B_m \}_{PB}= in\delta_{n+m,0}, \quad \{A_n,B_m \}_{PB} = 0.
\end{equation}
Using this oscillator algebra, it is straightforward to show that the charges satisfy one copy of Virasoro algebra. To see this, recall that the solutions of the conformal Killing equations \eqref{cct} on flat Carrollian backgrounds is given by 
\begin{equation}
	\xi^\tau=f'(\sigma)\tau +g(\sigma) \quad,\quad \xi^\sigma=f(\sigma),
\end{equation}
where $f$ and $g$ are functions of $\sigma$ only. Taking $f(\sigma)=\sum_{n}a_ne^{in\sigma}$, the charge in \eqref{charge} gives
\begin{eqnarray}
	Q_\xi = \sum_{n} a_n L_n; \quad L_n=\int d\sigma (\dot{\Phi}+\Phi')^2 e^{in\sigma}
\end{eqnarray}
Using the mode expansion, we get an expression for $L_n$ in terms of oscillators as
\begin{equation}
	L_n = \frac{1}{2}\sum_{p,q}\int d\sigma  A_{p}A_{q} e^{-i(p+q-n)\sigma} =\frac{1}{2}\sum_{p}A_pA_{n-p}.
\end{equation}
Alternately, we can also find expressions for $L_n$ and $M_n$ from the components of the improved stress tensor \eqref{improved-stress-tensor-mixed}. From $T_2 = 0$, using the expression $T_2 = \sum_n M_n e^{-in\sigma}$, we get $M_n = 0$. On the other hand,
\begin{equation}
	T_1 = \frac{1}{2}(\partial_{\sigma}\Phi + \partial_{\tau}\Phi)^2 = \sum_{m}\Big(\frac{1}{2}\sum_{n} A_{-n}A_{n+m}\Big)e^{-im\sigma}.
\end{equation}
Comparing this with the expression $T_1 = \sum_n (L_n - in\tau M_n)e^{-in\sigma}$, along with $M_n = 0$, one finds $L_m = \frac{1}{2}\sum_{n} A_{-n}A_{n+m}$. Then using the oscillator algebra \eqref{mixed-AB-algebra}, we find that these generators satisfy
\begin{equation}
	\{L_n, L_m\}_{PB} = -i(n-m) L_{n+m},
\end{equation}
which is one copy of the Virasoro algebra, thus providing a confirmation of what we argued from the Noether prescription as well. 

\subsection{Application: Flatspace Chiral Gravity}
To wrap up this section, let us comment on a possible intriguing connection of our mixed derivative theory with holography in asymptotically flat spacetimes. 

\medskip

As we mentioned in the introduction, the BMS$_3$ algebra is the asymptotic symmetry algebra of flat spacetimes at the null boundary. The details of the gravitational theory with asymptotically flat boundary conditions is encoded in the two central terms $c_L$ and $c_M$. For Einstein gravity, these turn out to be $c_L=0, c_M= 3/G$, where $G$ is the Newton's constant in $3d$. 

\medskip
If one wishes to add a non-zero central term, an easy way of doing this is to turn on a Gravitational Chern-Simons (GCS) term in the bulk theory, which now is given by the action{\footnote{Below the subscript EH means Einstein-Hilbert.}:
\be{}
S_{\text{bulk}} = S_{\text{EH}} + S_{\text{GCS}} = \frac{1}{16\pi G} \int d^3x \sqrt{-g} \ R + \frac{1}{32\pi G\mu} \int d^3x \sqrt{-g}\left(\Gamma \p \Gamma + \frac{2}{3} \Gamma^3\right). 
\ee
In the above, the indices on the connections have been suppressed ($\Gamma^a_{\ bc} \equiv \Gamma$). This theory goes under the name of Topologically Massive Gravity (TMG) \cite{Deser:1982vy}, and in the AdS context this has been used to provide evidence for a holographic theory of chiral gravity \cite{Li:2008dq} and logarithmic gravity \cite{Grumiller:2008qz}, when the parameter $\mu$ is tuned to a particular critical value. The central charges, in the case of asymptotically flat boundary conditions, become 
\be{}
c^{\text{(TMG)}}_L = \frac{3}{\mu G}, \quad c^{\text{(TMG)}}_M= \frac{3}{G}.
\ee
One can now take a particular interesting double scaling limit on this theory such that
\be{}
\mu = \e \to 0, \, G = \frac{1}{8k\e} \to \infty, \, \text{with} \quad \mu G = \frac{1}{8k}.
\ee
This limit send the Einstein-Hilbert term in the bulk action to zero and one is only left with the GCS term. This theory is called (conformal) Chern-Simons Gravity. 
\be{}
S_{\text{CSG}} = \frac{k}{4\pi} \int d^3x \sqrt{-g}\left(\Gamma \p \Gamma + \frac{2}{3} \Gamma^3\right). 
\ee
The central charge of the BMS$_3$ that is obtained as the asymptotic symmetries of this theory has 
\be{}
c^{\text{(CSG)}}_L = 24k, \quad c^{\text{(CSG)}}_M= 0.
\ee
It can further be shown that all the $M_n$ charges of the theory vanish onshell. The symmetries are thus reduced from the BMS$_3$ to a single copy of the Virasoro algebra. This is a feature of $3d$ BMS invariant field theories with $c_M=0$ and can be shown by an analysis of null vectors.{\footnote{Some evidence to the contrary have been reported recently in \cite{Hao:2021urq}. This depends crucially on the Jordan-block structures that arise in more general BMS invariant theories. It is possible that like the AdS story relating Chiral gravity and Log gravity \cite{Maloney:2009ck}, that Flatspace Chiral gravity exists as a sector (singlet sector) within a more general theory.}} The dual theory to CSG with asymptotically flat boundary conditions is thus a chiral $2d$ CFT governed by the symmetries of a single Virasoro algebra. This has been called Flatspace Chiral Gravity (F$\chi$G) \cite{Bagchi:2012yk}. See \cite{Bagchi:2018ryy} for a supersymmetric version. 

\medskip

Our discussions in this section have been focused on the mixed-derivative theory, which was defined on a null surface, but had a chiral Virasoro symmetry on-shell. We saw the truncation of symmetries from BMS$_3$ to Virasoro in this explicit example. It should be clear from the above discussion of F$\chi$G that the mixed derivative theory has features that resemble a possible dual field theory to Chern-Simons Gravity with flat boundary conditions. The theory lives on a null manifold, as is expected for the dual to CSG which should live on $\mathscr{I}^\pm$. And crucially, the symmetries of the theory reduce from BMS to a single copy of the Virasoro algebra. 

\bigskip

\section{Null string theories and Carrollian actions}\label{sec:ilst}
Tensionless or null string theories have been a well-studied example of a Carrollian limit of a relativistic theory, in this case that from the relativistic string worldsheet. The main ingredient in this theory has been the ILST action  \cite{Isberg:1993av} which replaces the well known Polyakov action on a Carrollian worldsheet,

\begin{equation}
	S_{\text{ILST}} = \int d^2\sigma V^{\mu}V^{\nu} \partial_{\mu}\Phi \partial_{\nu}\Phi, 
	\end{equation}
where $V^\mu$'s are vector densities under worldsheet diffeomorphisms and the equations of motion for them imply that the worldsheet metric has degenerated. Here we have suppressed the spacetime indices on the scalar field $\Phi$, and we assume a $D$ dimensional flat target space geometry. The above action can be thought of as intrinsic worldsheet action for null strings, or equivalently a UR limit of the tensile string theory \cite{Bagchi:2015nca}. In this section, we will revisit this action in light of our discussions throughout the bulk of this paper, and offer more insight into certain classes of actions which appeared in previous sections. 

\subsection{Mapping ILST and zweibein formulations from UR limit}\label{app:ILST-zweibeins-comparision}
As one can understand, taking a $T\to 0$ limit on a tensile sigma model action isn't well defined as the string tension appears as the coupling constant. Then one has to go to the Hamiltonian formulation and introduce auxiliary fields to impose string constraints. 
In the formulation of ILST, the worldsheet metric of a relativistic string (see \cite{Isberg:1993av} for more details) is parametrized by these two Lagrange multipliers $(\lambda,\rho)$ as
\begin{equation}\label{ILST-metric}
	g^{\mu\nu} = \begin{pmatrix}
		-1 & \rho \\
		\rho & \ -\rho^2 + 4\lambda^2 T^2
	\end{pmatrix}; \quad \det(g^{\mu\nu}) = -4\lambda^2 T^2.
\end{equation}
In the tensionless limit $T\rightarrow 0$, we see that the worldsheet (inverse) metric degenerates with $\det(g^{\mu\nu}) = 0$. For comparison with the zweibein formulation in previous sections, where we have $\det(g_{\mu\nu}) = 0$, we use the conformal invariance of the Polyakov action and perform a conformal transformation on \eqref{ILST-metric}:
\begin{equation}\label{lowertless}
	g_{\mu\nu} \rightarrow G_{\mu\nu} = \frac{T^2}{4}g_{\mu\nu} = \frac{1}{16\lambda^2}
	\begin{pmatrix}
		\rho^2 - 4\lambda^2 T^2 & \rho \\
		\rho & 1
	\end{pmatrix}; \quad \det(G_{\mu\nu}) = -\frac{T^2}{64\lambda^2},
\end{equation}
for which we have $\det(G_{\mu\nu}) = 0$ as $T\rightarrow 0$. One should note here that in general the conformal transformation we made stops making sense at $T=0$, and as a consequence the object $G_{\mu\nu}$ does not really exist in the ILST formalism.
However, continuing with the above, we can expand the metric density around $T=0$ as
\begin{equation}\label{ILST-metric-density}
	-\frac{T}{2}\sqrt{-G}G^{\mu\nu} = \frac{1}{4\lambda}
	\begin{pmatrix}
		1 & -\rho \\
		-\rho & \rho^2
	\end{pmatrix}
	-T^2 \begin{pmatrix}
		0 & 0 \\
		0 & \lambda
	\end{pmatrix}.
\end{equation}

\medskip

\noindent In the frame formulation, Carroll geometry is described by zweibeins $e^0_{\mu}$, $e^1_{\mu}$ which are obtained from an ultrarelativistic limit of the relativistic zweibeins as
\begin{equation}
	E^0_{\mu} = \epsilon e^0_{\mu}, \quad E^1_{\mu} = e^1_{\mu}, \quad E = \epsilon e; \quad \epsilon\rightarrow 0.
\end{equation}
Then expanding $G^{\mu\nu} = \eta^{AB}E^{\mu}_A E^{\nu}_B$, with $\sqrt{-G}=\sqrt{-\det(G_{\mu\nu})} = E$ and $\eta_{AB} = diag(-1,1)$, in $\epsilon$ we can write the metric density in terms of zweibeins as
\begin{eqnarray}\label{zweibeins-metric-density}
	-\epsilon\sqrt{-G}G^{\mu\nu} = e e^{\mu}_0 e^{\nu}_0 -\epsilon^2 e e^{\mu}_1 e^{\nu}_1.
\end{eqnarray}

\medskip

\noindent In the strict tensionless limit ($\e=0$, or $T=0$), we can write the degenerate metric density, \ie\ the first term in \eqref{ILST-metric-density} in terms of a vector density $V^{\mu}$:
\begin{equation}
	\lim_{T\rightarrow 0} -\frac{T}{2}\sqrt{-G}G^{\mu\nu} = V^{\mu}V^{\nu}; \quad V^{\mu} = \frac{1}{2\sqrt{\lambda}}(1,-\rho). \label{ILST-metric-density-V}
\end{equation}
In the UR limit, the first term in \eqref{zweibeins-metric-density} gives the degenerate metric density in terms of zweibeins:
\begin{equation}
	\lim_{\epsilon\rightarrow 0} -\epsilon\sqrt{-G}G^{\mu\nu} = e e^{\mu}_0 e^{\nu}_0. \label{zweibeins-metric-density-e}
\end{equation}
Identifying $\frac{T}{2}=\epsilon$,\footnote{On the flat (Minkowski) worldsheet described by $4\lambda=1$, $\rho=0$ giving $ds^2= -d\tau^2 + \frac{4}{T^2}d\sigma^2$, $ds^2=0$ gives $\vert\frac{d\sigma}{d\tau}\vert$=$\frac{T}{2}\sim$ speed of light on the worldsheet.} we see that the tensionless limit is equivalent to the UR limit. Then comparing \eqref{ILST-metric-density-V} and \eqref{zweibeins-metric-density-e}, we get
\begin{equation}
	V^{\mu} = \sqrt{e} e^{\mu}_0.
\end{equation}
So that the ILST action explicitly becomes our timelike action, and no analog of spacelike or mixed-derivative action can be found in this formalism. 
Further, one can naively compare the subleading terms in \eqref{ILST-metric-density} and \eqref{zweibeins-metric-density}, to get
\begin{equation}
	e (e^\t_1)^2 = 0, \quad e e^\t_1 e^\s_1 = 0, \quad e (e^\s_1)^2 = 4\lambda,
\end{equation}
so that we seem to conclude that $e^\t_1 = 0$ for Carroll geometry described in ILST formulation. But considering $G_{\mu\nu}|_{T=0}$ as the degenerate metric $h_{\mu\nu}$ in the Carrollian sense, we can see this conclusion may not be right as one cannot constrain $e^\t_1$ from this structure. In Appendix \ref{ApA}, we try to extend the ILST formalism beyond only the timelike action using this choice of $h_{\mu\nu}$. This is also important since $(\lambda,\rho)$ in general could be arbitrary functions of the coordinates, and the conditions of flat Carroll geometry should impose constraints on their structure. It does turn out that Carrollian structures impose that the Lagrange multipliers have to be constants, however an unfixed vielbien can freely exist in the theory as before.

\subsection{Different gauges}

Learning from the null string with degenerate worldsheet metric, \ie\ $\det(g_{\mu\nu}) = 0$, we now try to understand some of the actions we have discussed in this work. We would restrict ourselves for only one scalar for the time being and start with the theory described by the action \cite{Isberg:1993av}, which in general reads  
\begin{equation}
	S = \int d^2\sigma V^{\mu}V^{\nu} \partial_{\mu}\Phi \partial_{\nu}\Phi = \int d^2\sigma \frac{1}{4\lambda}\big(\dot{\Phi} - \rho\Phi'\big)^2, \label{ILST-action-2nd}
\end{equation}
but here $V^{\mu} = \frac{1}{2\sqrt{\lambda}}(1,-\rho)$ and $\lambda$, $\rho$ are generically non-zero constant Lagrange multipliers imposing the constraints
\begin{equation}
	P^2 = 0, \quad P \partial_\sigma \Phi = 0 \label{ILST-constraints-1}
\end{equation}
respectively. The stress tensor is given by varying the above action with respect to $V^\mu$:
\begin{equation}
	T^{\mu}_{\ \nu} = V^{\mu}V^{\alpha}\partial_{\alpha}\Phi\partial_{\nu}\Phi - \frac{\delta^{\mu}_{\nu}}{2}V^{\alpha}V^{\beta}\partial_{\alpha}\Phi\partial_{\beta}\Phi,
\end{equation}
whose components are
\begin{eqnarray}
	&& T^{\tau}_{\ \tau} = - T^{\sigma}_{\ \sigma} = \frac{1}{8\lambda}(\dot{\Phi}-\rho\Phi')(\dot{\Phi}+\rho\Phi'), \nonumber \\
	&& T^{\tau}_{\ \sigma} = \frac{1}{4\lambda}(\dot{\Phi}-\rho\Phi')\Phi', \quad T^{\sigma}_{\ \tau} = \frac{1}{4\lambda}(\dot{\Phi}-\rho\Phi')(-\rho\dot{\Phi}). \label{ILST-stress-tensor-2nd}
\end{eqnarray}
Using the expression for the conjugate momentum, $P = \frac{1}{2\lambda}(\dot{\Phi}-\rho\Phi')$, we can write the constraints \eqref{ILST-constraints-1} as
\begin{equation}
	\frac{1}{4\lambda^2}(\dot{\Phi}-\rho\Phi')^2 = 0, \quad \frac{1}{2\lambda}(\dot{\Phi}-\rho\Phi')\Phi' = 0. \label{ILST-constraints-2}
\end{equation}

\medskip

Equivalently, in first order formulation, the null string action, including the constraints, is written as
\begin{equation}
	S = \int d^2\sigma (P\dot{\Phi} - \lambda P^2 - \rho P\Phi'), \label{ILST-action-1st}
\end{equation}
with Hamiltonian equations of motions having the compact form
\be{}
V^\mu \partial_\mu \Phi = \lambda P,~~V^\mu \partial_\mu P = 0. 
\ee
The first order formalism is more interesting to use here as we can deal with a choice where $\lambda = 0$. Note here, due to structure of $V^\mu$ \eqref{ILST-metric-density-V} and equivalently that of $\t^\mu$, the $\lambda = 0$ dynamics cannot be reflected into the second order equations of motion and constraints. Let us discuss some explicitly chosen values of $(\lambda, \rho)$ to elaborate on this.

\subsubsection*{1) Static gauge: $\rho = 0$, $\lambda=0$}
The first order equations of motion here are 
\be{}
\dot\Phi = 0,~~\dot{P} = 0,
\ee
i.e. the solutions are static, there is no time evolution of the null string,
\be{}
\Phi(\t,\s)= \Phi_0(\s),~~P(\t,\s) = P_0(\s).
\ee
Using a periodic boundary condition in $\s$, we can expand the fields into modes
\be{}
\Phi_0(\s) = \sum_{n}A_n e^{-in\s},~~P_0(\s)=\sum_{n}P_n e^{-in\s}.
\ee
And the Poisson brackets of modes read
\begin{equation}
	\{A_n,P_m\}=\delta_{n+m,0}, \quad \{A_n,A_m\}=0, \quad \{P_n,P_m\}=0.
\end{equation}
Then ignoring constant factors, one could write the constraints \eqref{ILST-constraints-1} as
\be{}
P\Phi' \sim \sum_{n}L_n e^{-in\s}  = 0,~~~P^2 \sim \sum_{n}M_n e^{-in\s}  = 0,
\ee
where $L_n$ and $M_n$ again satisfy the classical part of the BMS$_3$ algebra. 

\subsubsection*{2) ``Timelike" gauge: $\rho = 0$, $4\lambda=1$}
From \eqref{ILST-action-2nd} we can see the second order action in this case is explicitly our timelike action 
\be{}
 S = \int d\t d\s (\partial_{\t}\Phi)^2.
\ee
The equations of motion in this case are 
\be{}
\dot\Phi = \frac{P}{4},~~\dot{P} = 0,
\ee
so that the solution with periodic boundary conditons has a linear dependence on time,
\be{}
\Phi(\t,\s) = \Phi_0(\s)+ \frac{\t}{4}P_0(\s).
\ee
One can again perform a mode expansion as before and compare this with the solutions for our timelike action \eqref{tlikesol}. This makes the structures of null string constraints evident: 
\be{}
P\Phi' \sim \sum_{n}(L_n-in\t M_n) e^{-in\s}  = 0,~~~P^2 \sim \sum_{n}M_n e^{-in\s}  = 0.
\ee
The generation of BMS$_3$ algebra is guaranteed from implementation of these constraints as well, due to the following intriguing automorphism of the algebra:
\be{}
L_n' \to L_n-in\t M_n,
\ee
which keeps the bracket structure invariant.

\subsubsection*{3) Ambitwistor gauge: $\rho = -1$, $\lambda=0$}

In second order formulation, the ambitwistor gauge is singular due to the presence of $\frac{1}{\lambda}$ in the action \eqref{ILST-action-2nd}. However in the first order formulation with the action \eqref{ILST-action-1st}, the ambitwistor gauge is well defined \cite{Casali:2016atr}. We should note herein lies the difference between our spacelike action with $e^\t_1 = 1$ \eqref{sp-action-et1-fixed} and pure Ambitwistor strings. In the Ambitwistor gauge, the Hamiltonian action \eqref{ILST-action-1st} becomes
\begin{equation}
	S = \int d^2\sigma (P\dot{\Phi} + P\Phi') = \int d^2\sigma P\partial_+\Phi.
\end{equation}
For $\lambda=0$, equations of motion become 
\begin{equation}
	\partial_+ P = 0, \quad \partial_+\Phi = 0; 
\end{equation}
giving rise to solutions and mode expansions of the form 
\be{}
\Phi = A(\sigma^-)= \sum_{n}A_n e^{-in\s^{-}},~~P = P(\s^-) = \sum_{n}P_n e^{-in\s^{-}}.
\ee
One could see these solutions do not suffer from a periodicity issue like we had in section \eqref{periodi} as there are no linear terms present in the mode expansion. Like our static gauge case, the constraints are simple 
\be{}
P\Phi' \sim \sum_{n}L_n e^{-in\s^{-}}  = 0,~~~P^2 \sim \sum_{n}M_n e^{-in\s^{-}}  = 0.
\ee
It goes without saying the these still generate the BMS$_3$ algebra.

\subsubsection*{4) ``Spacelike" gauge: $\rho = -1$, $4\lambda=1$}
This is explicitly the case we have discussed in  section \ref{sec:spacelike}.
In this gauge, the second order action \eqref{ILST-action-2nd} becomes
\begin{equation}
	S = \int d^2\sigma (\dot{\Phi} + \Phi')^2 = 4\int d\sigma^+ d\sigma^-(\partial_+\Phi)^2.
\end{equation}
The Hamilton's equation of motion and their solution turns out to be
\begin{equation}
	\partial_+ P = 0, \quad \partial_+\Phi = \lambda P; \qquad \Phi = A(\sigma^-) + \lambda \sigma^+ P(\sigma^-),
\end{equation}
which clearly lacks periodicity in $\s$, as we encountered before. 
The conjugate momentum is $P = 2(\dot{\Phi}+\Phi') = 4\partial_+\Phi$ and the constraints \eqref{ILST-constraints-2} become
\begin{equation}
	(\dot{\Phi}+\Phi')^2 = 0, \quad (\dot{\Phi}+\Phi')\Phi' = 0.
\end{equation}
The components of the stress tensor in first order formulation are
\begin{eqnarray}
	&& T^{\tau}_{\ \tau} = - T^{\sigma}_{\ \sigma} = \frac{P^2}{8} -\frac{P\Phi'}{2}, \quad T^{\tau}_{\ \sigma} = \frac{1}{2}P\Phi', \quad T^{\sigma}_{\ \tau} = \frac{P^2}{4} -\frac{P\Phi'}{2}.
\end{eqnarray}
We see that the action and the stress tensor above for the null string in the gauge $\rho=-1$, $4\lambda=1$ match exactly with those of the spacelike theory in \eqref{sp-et1fixed-stress-tensor-1st}. As we saw in that case, the symmetry structures are more evident in the ``lightcone'' coordinates, where it still generates a BMS$_3$ algebra. 

\medskip

The ``take home" message from this section is that the ILST formalism can generate certain gauge fixed versions of the geometrically obtained Carroll Conformal actions in the preceding sections. Specifically those are the purely timelike action and the spacelike action with a fixed  $e_1^\t =1$. However, there is no clear description for any cousin of the mixed derivative action in this formalism, making our discussion of the same even more unique. See Appendix \ref{ApB} for more discussion on this issue. The description of the mixed action as a null string theory, or a close relative thereof, thus remains an open question. We make a few speculative remarks about this at the end of our discussions section, which is up next.

\bigskip

\section{Discussions and conclusions}\label{sec:discussion}

\textbf{Summary}

\smallskip

In this paper, we have analyzed various classical properties of Carroll covariant actions for a massless scalar field in $2d$ Carroll spacetimes. Using the zweibeins $(\tau^{\mu},e^{\mu})$ describing the Carroll geometry to contract with $\partial_{\mu}\Phi$, we constructed three actions: the timelike action, the spacelike action and the mixed-derivative action. On a generic Carroll spacetime, each of these actions is Carroll diffeomorphism covariant and BMS-Weyl invariant. We saw that in our formalism, fixing the background to be flat Carroll spacetime leaves the component $e^\t$ arbitrary. From a purely geometric perspective, this unfixed zweibein $e^\t$ is truely redundant and leaving it arbitrary or choosing any constant value doesn't affect geometric quantities. However, on coupling the scalar field to the flat Carroll spacetime, $e^\t$ behaves non-trivially. In particular, its arbitrariness ensure BMS$_3$ invariance of all three actions for the scalar field on the flat Carroll spacetime. But evidently, on taking $e^\t$ to be constant, the resultant spacelike and mixed-derivative actions have reduced set of (offshell) residual symmetries, \ie\ only translations and scaling. This leads us to treat the spacelike and mixed-derivative actions differently.

\medskip

For the spacelike action with $e^\t$ fixed to a constant value, we found that we recover the offshell BMS$_3$ symmetries by transforming to lightcone coordinates. We would like to emphasize that this is not inconsistent with the earlier statement about residual symmetries being reduced to only translations and scaling. To clarify, this reduction of residual symmetries happens in the standard (flat) Carroll frame parametrized by ($\tau,\sigma$) coordinates, whereas the BMS$_3$ symmetries are recovered in lightcone coordinates which parametrize an inequivalent (flat) Carroll frame (see sec.~\ref{sec:2d-flat-Carroll} for more details).

\medskip

Our novel mixed-derivative theory turned out to have quite a rich structure. Offshell, the mixed-derivative action can be argued to possess a gauge symmetry under gauge transformations of $e^\t$. This gauge symmetry could be interpreted to have a compensatory effect that allows us to gauge-fix $e^\t$ to a constant value without sacrificing the BMS symmetries. Onshell, the mixed-derivative theory shows enhancement in the symmetries to one copy of the Virasoro algebra.

\medskip

To close our discussion, we made contact with existing literature on Tensionless strings and described how it fits as a subset of the classes of BMS invariant actions we have discussed in this work. Thus this work encompassed examples of BMS as both global and gauge symmetries in physical systems. 

\bigskip

\textbf{Future directions}

\smallskip
This paper has been a small first step along a very interesting thoroughfare, and we certainly have miles to go ahead. Since we focussed solely on the $2d$ case here, the immediate extension that comes to mind is to discuss analogous actions in higher dimensions. A nice discussion has already appeared in \cite{Gupta:2020dtl}, but the corresponding analogue of our mixed derivative action remains absent in that formalism. 
\medskip

Another point of interest would be to geometrically understand the significance of the unfixed vielbien $e^\t$. Although we could show there is a gauge redundancy of that quantity in our formalism, the actual nature of the freedom associated to it on a null manifold is subtle. Especially in the case of $3d$ conformal Carroll actions, which should be important in studying duals of $4d$ flat space gravity, this gauge redundancy may hold some more surprises in store. Similar questions actually would survive if we want to interpret our two dimensional actions with an unfixed $e^\t$ as valid string theories. We plan to come back to these problems in an upcoming work.
\medskip

A very natural follow-up work would be to try and quantize the theories we explored in this paper. Although for the timelike case, a large amount of literature exist on quantum structures \cite{Bagchi:2020fpr,Bagchi:2021rfw}, and the vacuum structure associated to Ambitwistor string is also very well explored  \cite{Casali:2016atr, Casali:2017zkz, Bagchi:2020fpr}, the quantization associated to the mixed-derivative action demands immediate attention. As this is a manifestly chiral theory on shell, it begs the question: whether it corresponds to a chiral half of a $2d$ CFT, or is there even more to it? To start with, determination of central charges for the chiral algebra will be an important step here. It will also be instructive to find quantized Carroll theories beyond covariant phase space methods, perhaps using a path integral formalism. However the degenerate nature inherent to these theories will require a major overhaul for the known techniques. 

\medskip

One of our principal motivations for the investigation of Carrollian structures is to establish some version of Flat Holography in different dimensions. To this end, we would like to construct full supersymmetric Carroll CFTs in future endeavours. Supersymmetric Carrollian theories have been explored in various contexts, like supersymmetric null strings \cite{Bagchi:2016yyf,Bagchi:2017cte,Bagchi:2018wsn}, Flat Supergravities \cite{Lodato:2016alv, Bagchi:2018ryy}, and very recently in the form of Carrollian $\mathcal{N} = 1$ superconformal theory \cite{Bagchi:2022owq}. In all of these cases it has been firmly established that adding Carrollian spinors make the structure of these theories extremely rich. Just to take the example in $2d$, it has been shown that Super-BMS$_3$ algebra has two distinct avatars, called the homogeneous and inhomogeneous (or democratic and despotic) supersymmetric theories, based on different inequivalent representation of the degenerate Clifford algebra. A thorough analysis of the Spin Group defined on the intrinsically Carrollian metric data is still an open problem, and a genuinely intricate one at that. We should be able to clarify these structures further in the near future. 
 
\medskip

Before we end, a final word about our new mixed derivative theory and possible connections to null strings. One of the problematic issues in recent attempts to construct a theory of tensionless strings is the lack of understanding of open strings. There have been attempts at this earlier \cite{Bonelli:2003kh, Lindstrom:2003mg}, but the manifest null structures that should be associated with such a construction are far from obvious. To emphasise this point, we note that \cite{Casali:2017zkz} claims the open null string algebras found by contraction (e.g. in \cite{Bonelli:2003kh}) cannot be found by imposing appropriate boundary conditions on the null string. The $2d$ open string worldsheet has to become null and that would mean the emergence of a BMS type algebra. The fact that our mixed derivative theory starts off having BMS invariance and then reduces to a single copy of Virasoro onshell is rather tantalising and perhaps indicative of tensionless open strings. It would be instructive to follow this lead further and also figure out whether this can have further implications to the very interesting closed-to-open transitions discovered in \cite{Bagchi:2019cay}.

\bigskip

\bigskip

\section*{Acknowledgements}

\medskip

AB is partially supported by a Swarnajayanti fellowship of the Department of Science and Technology, India and by the following grants from the Science and Engineering Research Board: SB/SJF/2019-20/08, MTR/2017/000740, CGR/2020/002035. SD is supported by grant number
09/092(0971)/2017-EMR-I from Council of Scientific and Industrial Research (CSIR).
The work of ArB is supported by the Quantum Gravity Unit of the Okinawa Institute of Science and Technology Graduate University (OIST). He would also like to thank Physics Department, Kyoto University for hospitality during the final part of this work.

\bigskip
\bigskip
\bigskip

\section*{APPENDICES}

\appendix

\section{ILST action vs. vielbein formalism} \label{ApA}
Inspired by the degenerate metric that appears in \eqref{lowertless} at the tensionless point $T=0$,
we consider the degenerate spacetime metric $h_{\mu\nu} = e_{\mu}e_{\nu}$ with $e_t$ arbitrary:
\begin{equation}
	h_{\mu\nu} = \Big(\begin{matrix}
		e_\t^2 & e_\t \\
		e_\t & 1
	\end{matrix}\Big); \quad \det(h_{\mu\nu}) = 0.
\end{equation}
The projective inverse relations \eqref{zweibeins-inverse} give
\begin{eqnarray}
	&& e^\t e_\t + e^\s = 1, \quad e_\t\tau^\t + \tau^\s = 0, \quad e^\t\tau_\t + e^\s\tau_\s = 0, \quad \tau^\t\tau_\t + \tau^\s\tau_\s = 1, \nonumber \\
	&& e_\t e^\t + \tau_\t\tau^\t = 1, \quad e^\s + \tau_\s\tau^\s = 1, \quad e_\t e^\s + \tau_\t\tau^\s = 0, \quad e^\t +\tau_\s\tau^\t = 0.
\end{eqnarray}
Note that $e_\t = 0$ corresponds to the standard flat Carroll spacetime \eqref{flat-Carroll-metric}. So here we want to take $e_\t\neq 0$. Then from $e_\t\tau^t + \tau^\s = 0$, we see that either $\tau^\s = 0 = \tau^\t$ or $\tau^\s \neq 0 \neq \tau^\t$. Since $\tau^{\mu}$ is no-where vanishing, we take $\tau^\s \neq 0 \neq \tau^\t$. Then the ILST parametrization of the null metric with $V^{\mu} = \frac{1}{2\sqrt{\lambda}}(1,-\rho)$, where $\lambda$ and $\rho$ are arbitrary functions on the Carroll spacetime, corresponds to
\begin{equation}
	V^{\mu} = \sqrt{e}\tau^{\mu} \qquad \implies\quad e = 4\lambda, \quad \tau^\t = \frac{1}{4\lambda}, \quad \tau^\s = -\frac{\rho}{4\lambda}.
\end{equation}
We note that the determinant $e\neq 0$ and a smooth $\tau^{\mu}$ requires $\lambda\neq 0,\infty$. Then from the projective inverse relations above, we have the components of zweibeins as
\begin{eqnarray}
	&& \tau^\t = \frac{1}{4\lambda}, \quad \tau^\s = -\frac{\rho}{4\lambda}, \quad e^\t = -\frac{\tau_\s}{4\lambda}, \quad e^\s = 1 + \frac{\rho\tau_\s}{4\lambda}, \nonumber \\
	&& \tau_\t = 4\lambda + \rho\tau_\s, \quad \tau_\s \sim \text{arbitrary}, \quad e_\t = \rho, \quad e_\s = 1, \label{ILST-zweibeins}
\end{eqnarray}
where $\rho$, $\lambda$ and $\tau_\s$ are arbitrary functions as of now.

For \eqref{ILST-zweibeins} to describe a strong flat Carroll spacetime, we require the spacetime connection \eqref{2d-affine-connection} to vanish. Let us write the components of the spacetime connection for \eqref{ILST-zweibeins}:
\begin{eqnarray}
	&& \Gamma^\t_{\t\t} = \frac{\partial_\t\lambda}{\lambda}, \quad \Gamma^\t_{\s\t} = \frac{\partial_\s\lambda}{\lambda}, \quad \Gamma^\s_{\t\t} = \partial_\t\rho -\frac{\rho}{\lambda}\partial_\t\lambda, \quad \Gamma^\s_{\s\t} = \partial_\s\rho -\frac{\rho}{\lambda}\partial_\s\lambda, \nonumber \\
	&& \Gamma^\t_{\t\s} = \frac{1}{4\lambda}(\partial_\t\tau_\s - \Omega_\t), \quad \Gamma^\t_{\s\s} = \frac{1}{4\lambda}(\partial_\s\tau_\s - \Omega_\s), \nonumber \\
	&& \Gamma^\s_{\t\s} = -\frac{\rho}{4\lambda}(\partial_\t\tau_\s - \Omega_\t), \quad \Gamma^\s_{\s\s} = -\frac{\rho}{4\lambda}(\partial_\s\tau_\s - \Omega_\s).
\end{eqnarray}
We see that $\Gamma^{\rho}_{\mu\nu} = 0$ restricts the arbitrary functions in \eqref{ILST-zweibeins} and determines the boost-spin connection as
\begin{equation}
	\lambda = \text{constant}, \quad \rho = \text{constant}, \quad \tau_\s \sim \text{arbitrary}; \quad \Omega_{\mu} = \partial_{\mu}\tau_\s.
\end{equation}
We can easily see that the boost curvature and the Riemann tensor vanishes with these choices. We conclude that the ILST parametrization $V^{\mu}$ describes a strong flat Carroll spacetime only for constant $\lambda$ and $\rho$. In that case, the (BMS) conformal Killing equations become
\begin{equation}
	\partial_t\xi^\t - \partial_\s\xi^\s - 2\rho\partial_\s\xi^\t = 0, \quad \partial_\t\xi^\s + \rho^2\partial_\s\xi^\t = 0.
\end{equation}
Putting $\rho = -1$ one can generate the killing equations mentioned in \eqref{lc-BMS-in-tx-1}.

\section{Parameter space for ILST actions}\label{ApB}
As we discussed earlier, there are two generic constraints for a null string, both coming from dynamics of the scalar field $\Phi$, and can be written as
\be{}
P^2 = 0, ~~ P\cdot \p_\s \Phi =0,
\ee
$P$ being the canonical momentum of the system. These are the constraints which are supposed to generate BMS$_3$ symmetries, with canonical brackets of $\{\Phi,\Pi\}$ as input. A generic action for the system is then
\begin{equation}
	S = \int d^2\sigma (P\dot{\Phi} + \lambda P^2 +\rho P\Phi'), \label{ILST-action-2nd1}
\end{equation}
where $\lambda$ and $\rho$ are lagrange multipliers (compare with \eqref{ILST-action-1st}). For a second order dynamical system, we can fix a gauge where $P$ is generic linear in the derivatives of the field $\Phi$,
\be{}
P = \a \dot{\Phi}+\b \Phi'
\ee
Putting this in the action and using $P = \frac{\partial L}{\partial \dot{\Phi}}$, we can solve for $(\lambda, \rho)$ in terms of the gauge parameters $(\a,\b)$, which could in principle be general functions of $\phi$, however not of its derivatives. This leads us to,
\be{}
\lambda = -\frac{1}{2\a},~~\rho = \frac{\b}{\a}, ~~\a\neq0.
\ee
So the Lagrangian for the gauge fixed system becomes 
\be{}
\mathcal{L} = \frac{\a}{2}\dot{\Phi}^2+ \frac{\b^2}{2\a}\Phi'^2+\b \dot{\Phi}\cdot \Phi'.
\ee
Now $(\a,\b)$ remains generally undetermined parameters, but as long as we have the constraints in place, this should be a general BMS$_3$ invariant theory in two dimensions. Let us consider distinct cases:
\paragraph{Case 1: $\b =0$ and $\a =1$}:
We have the Timelike action of $\L_1 = \frac{\dot{\Phi}^2}{2}$.
\paragraph{Case 2: $\b =\pm 1$ and $\a =1$}
The action becomes chiral  $\L=\frac{1}{2} \left( \dot{\Phi}\pm \Phi'\right)^2$. This action is nothing but our spacelike action with fixed $e^\t =1$,
\be{}
\L_2 = \frac{1}{2}(\partial_{z}\Phi)^2,
\ee
written in holomorphic coordinates.
The Hamiltonian of such a theory takes the form 
\be{}
\mathcal{H} = \frac{1}{2}(\dot{\Phi}^2-\Phi'^2) = \frac{1}{2}\partial_z \Phi\partial_{\bar{z}}\Phi,
\ee
which incidentally is the lagrangian for a $2d$ CFT.
But there could be more self consistent branches of solutions for the gauge parameters $(\a,\b)$. For example if $\a =0$ then above solution is not valid and the unfixed lagrangian takes a form 
\be{}
\L = \b \dot{\Phi}\cdot \Phi'+(\lambda\b^2+\rho\b) \Phi'^2.
\ee
Now there is no way to trade in $(\lambda,\rho)$ in terms of $(\a,\b)$ anymore and $\b$ remains unfixed. Although the constraints are robust
\be{}
\Phi'^2 = 0, ~~\dot{\Phi}\cdot \Phi'=0.
\ee
A naive choice in the parameter space could be that of $\b=1$ and correspondingly, $\lambda\beta+\rho = \pm 1$, which leads to:
\be{}
\L_3 = \Phi'(\dot{\Phi}\pm \Phi'),
\ee
which is also the Floreanini-Jackiw action \cite{Floreanini:1987as}, \ie\ the time-space inverted version of \eqref{Sm-gaugefixed}.

\newpage

\bibliographystyle{JHEP}
\bibliography{ref}

\providecommand{\href}[2]{#2}\begingroup\raggedright\begin{thebibliography}{10}

\bibitem{LBLL}
J.~Levy-Leblond, \emph{{Une nouvelle limite non-relativiste du group de
  Poincare}}, {\emph{Ann.Inst.Henri Poincare} {\bfseries 3} (1965) 1}.

\bibitem{Henneaux:1979vn}
M.~Henneaux, \emph{{Geometry of Zero Signature Space-times}}, {\emph{Bull. Soc.
  Math. Belg.} {\bfseries 31} (1979) 47}.

\bibitem{Duval:2014uoa}
C.~Duval, G.~W. Gibbons, P.~A. Horvathy and P.~M. Zhang, \emph{{Carroll versus
  Newton and Galilei: two dual non-Einsteinian concepts of time}},
  \href{https://doi.org/10.1088/0264-9381/31/8/085016}{\emph{Class. Quant.
  Grav.} {\bfseries 31} (2014) 085016}
  [\href{https://arxiv.org/abs/1402.0657}{{\ttfamily 1402.0657}}].

\bibitem{Duval:2014uva}
C.~Duval, G.~W. Gibbons and P.~A. Horvathy, \emph{{Conformal Carroll groups and
  BMS symmetry}},
  \href{https://doi.org/10.1088/0264-9381/31/9/092001}{\emph{Class. Quant.
  Grav.} {\bfseries 31} (2014) 092001}
  [\href{https://arxiv.org/abs/1402.5894}{{\ttfamily 1402.5894}}].

\bibitem{Duval:2014lpa}
C.~Duval, G.~W. Gibbons and P.~A. Horvathy, \emph{{Conformal Carroll groups}},
  \href{https://doi.org/10.1088/1751-8113/47/33/335204}{\emph{J. Phys.}
  {\bfseries A47} (2014) 335204}
  [\href{https://arxiv.org/abs/1403.4213}{{\ttfamily 1403.4213}}].

\bibitem{Bondi:1}
H.~Bondi, M.~G.~J. van~der Burg and A.~W.~K. Metzner, \emph{{Gravitational
  waves in general relativity. 7. Waves from axisymmetric isolated systems}},
  \href{https://doi.org/10.1098/rspa.1962.0161}{\emph{Proc. Roy. Soc. Lond.}
  {\bfseries A269} (1962) 21}.

\bibitem{Sachs:1962zza}
R.~Sachs, \emph{{Asymptotic symmetries in gravitational theory}},
  \href{https://doi.org/10.1103/PhysRev.128.2851}{\emph{Phys. Rev.} {\bfseries
  128} (1962) 2851}.

\bibitem{Barnich:2006av}
G.~Barnich and G.~Compere, \emph{{Classical central extension for asymptotic
  symmetries at null infinity in three spacetime dimensions}},
  \href{https://doi.org/10.1088/0264-9381/24/5/F01,
  10.1088/0264-9381/24/11/C01}{\emph{Class. Quant. Grav.} {\bfseries 24} (2007)
  F15} [\href{https://arxiv.org/abs/gr-qc/0610130}{{\ttfamily gr-qc/0610130}}].

\bibitem{Bagchi:2012cy}
A.~Bagchi and R.~Fareghbal, \emph{{BMS/GCA Redux: Towards Flatspace Holography
  from Non-Relativistic Symmetries}},
  \href{https://doi.org/10.1007/JHEP10(2012)092}{\emph{JHEP} {\bfseries 10}
  (2012) 092} [\href{https://arxiv.org/abs/1203.5795}{{\ttfamily 1203.5795}}].

\bibitem{Bagchi:2009pe}
A.~Bagchi, R.~Gopakumar, I.~Mandal and A.~Miwa, \emph{{GCA in 2d}},
  \href{https://doi.org/10.1007/JHEP08(2010)004}{\emph{JHEP} {\bfseries 08}
  (2010) 004} [\href{https://arxiv.org/abs/0912.1090}{{\ttfamily 0912.1090}}].

\bibitem{Bagchi:2010zz}
A.~Bagchi, \emph{{Correspondence between Asymptotically Flat Spacetimes and
  Nonrelativistic Conformal Field Theories}},
  \href{https://doi.org/10.1103/PhysRevLett.105.171601}{\emph{Phys. Rev. Lett.}
  {\bfseries 105} (2010) 171601}
  [\href{https://arxiv.org/abs/1006.3354}{{\ttfamily 1006.3354}}].

\bibitem{Maldacena:1997re}
J.~M. Maldacena, \emph{{The Large N limit of superconformal field theories and
  supergravity}}, \href{https://doi.org/10.1023/A:1026654312961,
  10.4310/ATMP.1998.v2.n2.a1}{\emph{Int. J. Theor. Phys.} {\bfseries 38} (1999)
  1113} [\href{https://arxiv.org/abs/hep-th/9711200}{{\ttfamily
  hep-th/9711200}}].

\bibitem{Barnich:2010eb}
G.~Barnich and C.~Troessaert, \emph{{Aspects of the BMS/CFT correspondence}},
  \href{https://doi.org/10.1007/JHEP05(2010)062}{\emph{JHEP} {\bfseries 05}
  (2010) 062} [\href{https://arxiv.org/abs/1001.1541}{{\ttfamily 1001.1541}}].

\bibitem{Bagchi:2012yk}
A.~Bagchi, S.~Detournay and D.~Grumiller, \emph{{Flat-Space Chiral Gravity}},
  \href{https://doi.org/10.1103/PhysRevLett.109.151301}{\emph{Phys. Rev. Lett.}
  {\bfseries 109} (2012) 151301}
  [\href{https://arxiv.org/abs/1208.1658}{{\ttfamily 1208.1658}}].

\bibitem{Bagchi:2012xr}
A.~Bagchi, S.~Detournay, R.~Fareghbal and J.~Sim{\'o}n, \emph{{Holography of 3D
  Flat Cosmological Horizons}},
  \href{https://doi.org/10.1103/PhysRevLett.110.141302}{\emph{Phys. Rev. Lett.}
  {\bfseries 110} (2013) 141302}
  [\href{https://arxiv.org/abs/1208.4372}{{\ttfamily 1208.4372}}].

\bibitem{Barnich:2012xq}
G.~Barnich, \emph{{Entropy of three-dimensional asymptotically flat
  cosmological solutions}},
  \href{https://doi.org/10.1007/JHEP10(2012)095}{\emph{JHEP} {\bfseries 10}
  (2012) 095} [\href{https://arxiv.org/abs/1208.4371}{{\ttfamily 1208.4371}}].

\bibitem{Bagchi:2013qva}
A.~Bagchi and R.~Basu, \emph{{3D Flat Holography: Entropy and Logarithmic
  Corrections}}, \href{https://doi.org/10.1007/JHEP03(2014)020}{\emph{JHEP}
  {\bfseries 03} (2014) 020} [\href{https://arxiv.org/abs/1312.5748}{{\ttfamily
  1312.5748}}].

\bibitem{Bagchi:2015wna}
A.~Bagchi, D.~Grumiller and W.~Merbis, \emph{{Stress tensor correlators in
  three-dimensional gravity}},
  \href{https://doi.org/10.1103/PhysRevD.93.061502}{\emph{Phys. Rev. D}
  {\bfseries 93} (2016) 061502}
  [\href{https://arxiv.org/abs/1507.05620}{{\ttfamily 1507.05620}}].

\bibitem{Bagchi:2014iea}
A.~Bagchi, R.~Basu, D.~Grumiller and M.~Riegler, \emph{{Entanglement entropy in
  Galilean conformal field theories and flat holography}},
  \href{https://doi.org/10.1103/PhysRevLett.114.111602}{\emph{Phys. Rev. Lett.}
  {\bfseries 114} (2015) 111602}
  [\href{https://arxiv.org/abs/1410.4089}{{\ttfamily 1410.4089}}].

\bibitem{Jiang:2017ecm}
H.~Jiang, W.~Song and Q.~Wen, \emph{{Entanglement Entropy in Flat Holography}},
  \href{https://doi.org/10.1007/JHEP07(2017)142}{\emph{JHEP} {\bfseries 07}
  (2017) 142} [\href{https://arxiv.org/abs/1706.07552}{{\ttfamily
  1706.07552}}].

\bibitem{Hijano:2017eii}
E.~Hijano and C.~Rabideau, \emph{{Holographic entanglement and Poincar\'e
  blocks in three-dimensional flat space}},
  \href{https://doi.org/10.1007/JHEP05(2018)068}{\emph{JHEP} {\bfseries 05}
  (2018) 068} [\href{https://arxiv.org/abs/1712.07131}{{\ttfamily
  1712.07131}}].

\bibitem{Bagchi:2013lma}
A.~Bagchi, S.~Detournay, D.~Grumiller and J.~Simon, \emph{{Cosmic Evolution
  from Phase Transition of Three-Dimensional Flat Space}},
  \href{https://doi.org/10.1103/PhysRevLett.111.181301}{\emph{Phys. Rev. Lett.}
  {\bfseries 111} (2013) 181301}
  [\href{https://arxiv.org/abs/1305.2919}{{\ttfamily 1305.2919}}].

\bibitem{Detournay:2014fva}
S.~Detournay, D.~Grumiller, F.~Sch\"oller and J.~Sim\'on, \emph{{Variational
  principle and one-point functions in three-dimensional flat space Einstein
  gravity}}, \href{https://doi.org/10.1103/PhysRevD.89.084061}{\emph{Phys. Rev.
  D} {\bfseries 89} (2014) 084061}
  [\href{https://arxiv.org/abs/1402.3687}{{\ttfamily 1402.3687}}].

\bibitem{Afshar:2013vka}
H.~Afshar, A.~Bagchi, R.~Fareghbal, D.~Grumiller and J.~Rosseel, \emph{{Spin-3
  Gravity in Three-Dimensional Flat Space}},
  \href{https://doi.org/10.1103/PhysRevLett.111.121603}{\emph{Phys. Rev. Lett.}
  {\bfseries 111} (2013) 121603}
  [\href{https://arxiv.org/abs/1307.4768}{{\ttfamily 1307.4768}}].

\bibitem{Gonzalez:2013oaa}
H.~A. Gonzalez, J.~Matulich, M.~Pino and R.~Troncoso, \emph{{Asymptotically
  flat spacetimes in three-dimensional higher spin gravity}},
  \href{https://doi.org/10.1007/JHEP09(2013)016}{\emph{JHEP} {\bfseries 09}
  (2013) 016} [\href{https://arxiv.org/abs/1307.5651}{{\ttfamily 1307.5651}}].

\bibitem{Hartong:2015usd}
J.~Hartong, \emph{{Holographic Reconstruction of 3D Flat Space-Time}},
  \href{https://doi.org/10.1007/JHEP10(2016)104}{\emph{JHEP} {\bfseries 10}
  (2016) 104} [\href{https://arxiv.org/abs/1511.01387}{{\ttfamily
  1511.01387}}].

\bibitem{Hartong:2015xda}
J.~Hartong, \emph{{Gauging the Carroll Algebra and Ultra-Relativistic
  Gravity}}, \href{https://doi.org/10.1007/JHEP08(2015)069}{\emph{JHEP}
  {\bfseries 08} (2015) 069}
  [\href{https://arxiv.org/abs/1505.05011}{{\ttfamily 1505.05011}}].

\bibitem{Bagchi:2016geg}
A.~Bagchi, M.~Gary and Zodinmawia, \emph{{Bondi-Metzner-Sachs bootstrap}},
  \href{https://doi.org/10.1103/PhysRevD.96.025007}{\emph{Phys. Rev.}
  {\bfseries D96} (2017) 025007}
  [\href{https://arxiv.org/abs/1612.01730}{{\ttfamily 1612.01730}}].

\bibitem{Barnich:2014cwa}
G.~Barnich, L.~Donnay, J.~Matulich and R.~Troncoso, \emph{{Asymptotic
  symmetries and dynamics of three-dimensional flat supergravity}},
  \href{https://doi.org/10.1007/JHEP08(2014)071}{\emph{JHEP} {\bfseries 08}
  (2014) 071} [\href{https://arxiv.org/abs/1407.4275}{{\ttfamily 1407.4275}}].

\bibitem{Fareghbal:2014qga}
R.~Fareghbal and A.~Naseh, \emph{{Aspects of Flat/CCFT Correspondence}},
  \href{https://doi.org/10.1088/0264-9381/32/13/135013}{\emph{Class. Quant.
  Grav.} {\bfseries 32} (2015) 135013}
  [\href{https://arxiv.org/abs/1408.6932}{{\ttfamily 1408.6932}}].

\bibitem{Grumiller:2019xna}
D.~Grumiller, P.~Parekh and M.~Riegler, \emph{{Local quantum energy conditions
  in non-Lorentz-invariant quantum field theories}},
  \href{https://arxiv.org/abs/1907.06650}{{\ttfamily 1907.06650}}.

\bibitem{Ciambelli:2018wre}
L.~Ciambelli, C.~Marteau, A.~C. Petkou, P.~M. Petropoulos and K.~Siampos,
  \emph{{Flat holography and Carrollian fluids}},
  \href{https://doi.org/10.1007/JHEP07(2018)165}{\emph{JHEP} {\bfseries 07}
  (2018) 165} [\href{https://arxiv.org/abs/1802.06809}{{\ttfamily
  1802.06809}}].

\bibitem{Bagchi:2016bcd}
A.~Bagchi, R.~Basu, A.~Kakkar and A.~Mehra, \emph{{Flat Holography: Aspects of
  the dual field theory}},
  \href{https://doi.org/10.1007/JHEP12(2016)147}{\emph{JHEP} {\bfseries 12}
  (2016) 147} [\href{https://arxiv.org/abs/1609.06203}{{\ttfamily
  1609.06203}}].

\bibitem{Strominger:2017zoo}
A.~Strominger, \emph{{Lectures on the Infrared Structure of Gravity and Gauge
  Theory}},  \href{https://arxiv.org/abs/1703.05448}{{\ttfamily 1703.05448}}.

\bibitem{Pasterski:2021rjz}
S.~Pasterski, \emph{{Lectures on celestial amplitudes}},
  \href{https://doi.org/10.1140/epjc/s10052-021-09846-7}{\emph{Eur. Phys. J. C}
  {\bfseries 81} (2021) 1062}
  [\href{https://arxiv.org/abs/2108.04801}{{\ttfamily 2108.04801}}].

\bibitem{Raclariu:2021zjz}
A.-M. Raclariu, \emph{{Lectures on Celestial Holography}},
  \href{https://arxiv.org/abs/2107.02075}{{\ttfamily 2107.02075}}.

\bibitem{Donnay:2022aba}
L.~Donnay, A.~Fiorucci, Y.~Herfray and R.~Ruzziconi, \emph{{A Carrollian
  Perspective on Celestial Holography}},
  \href{https://arxiv.org/abs/2202.04702}{{\ttfamily 2202.04702}}.

\bibitem{Bagchi:2022emh}
A.~Bagchi, S.~Banerjee, R.~Basu and S.~Dutta, \emph{{Scattering Amplitudes:
  Celestial and Carrollian}},
  \href{https://arxiv.org/abs/2202.08438}{{\ttfamily 2202.08438}}.

\bibitem{Hawking:2016msc}
S.~W. Hawking, M.~J. Perry and A.~Strominger, \emph{{Soft Hair on Black
  Holes}}, \href{https://doi.org/10.1103/PhysRevLett.116.231301}{\emph{Phys.
  Rev. Lett.} {\bfseries 116} (2016) 231301}
  [\href{https://arxiv.org/abs/1601.00921}{{\ttfamily 1601.00921}}].

\bibitem{Hawking:2016sgy}
S.~W. Hawking, M.~J. Perry and A.~Strominger, \emph{{Superrotation Charge and
  Supertranslation Hair on Black Holes}},
  \href{https://doi.org/10.1007/JHEP05(2017)161}{\emph{JHEP} {\bfseries 05}
  (2017) 161} [\href{https://arxiv.org/abs/1611.09175}{{\ttfamily
  1611.09175}}].

\bibitem{Donnay:2015abr}
L.~Donnay, G.~Giribet, H.~A. Gonzalez and M.~Pino, \emph{{Supertranslations and
  Superrotations at the Black Hole Horizon}},
  \href{https://doi.org/10.1103/PhysRevLett.116.091101}{\emph{Phys. Rev. Lett.}
  {\bfseries 116} (2016) 091101}
  [\href{https://arxiv.org/abs/1511.08687}{{\ttfamily 1511.08687}}].

\bibitem{Afshar:2016wfy}
H.~Afshar, S.~Detournay, D.~Grumiller, W.~Merbis, A.~Perez, D.~Tempo et~al.,
  \emph{{Soft Heisenberg hair on black holes in three dimensions}},
  \href{https://doi.org/10.1103/PhysRevD.93.101503}{\emph{Phys. Rev. D}
  {\bfseries 93} (2016) 101503}
  [\href{https://arxiv.org/abs/1603.04824}{{\ttfamily 1603.04824}}].

\bibitem{Penna:2017bdn}
R.~F. Penna, \emph{{Near-horizon BMS symmetries as fluid symmetries}},
  \href{https://doi.org/10.1007/JHEP10(2017)049}{\emph{JHEP} {\bfseries 10}
  (2017) 049} [\href{https://arxiv.org/abs/1703.07382}{{\ttfamily
  1703.07382}}].

\bibitem{Donnay:2019jiz}
L.~Donnay and C.~Marteau, \emph{{Carrollian Physics at the Black Hole
  Horizon}}, \href{https://doi.org/10.1088/1361-6382/ab2fd5}{\emph{Class.
  Quant. Grav.} {\bfseries 36} (2019) 165002}
  [\href{https://arxiv.org/abs/1903.09654}{{\ttfamily 1903.09654}}].

\bibitem{Carlip:2017xne}
S.~Carlip, \emph{{Black Hole Entropy from Bondi-Metzner-Sachs Symmetry at the
  Horizon}}, \href{https://doi.org/10.1103/PhysRevLett.120.101301}{\emph{Phys.
  Rev. Lett.} {\bfseries 120} (2018) 101301}
  [\href{https://arxiv.org/abs/1702.04439}{{\ttfamily 1702.04439}}].

\bibitem{Carlip:2019dbu}
S.~Carlip, \emph{{Near-horizon Bondi-Metzner-Sachs symmetry, dimensional
  reduction, and black hole entropy}},
  \href{https://doi.org/10.1103/PhysRevD.101.046002}{\emph{Phys. Rev. D}
  {\bfseries 101} (2020) 046002}
  [\href{https://arxiv.org/abs/1910.01762}{{\ttfamily 1910.01762}}].

\bibitem{Schild:1976vq}
A.~Schild, \emph{{Classical Null Strings}},
  \href{https://doi.org/10.1103/PhysRevD.16.1722}{\emph{Phys. Rev.} {\bfseries
  D16} (1977) 1722}.

\bibitem{Isberg:1993av}
J.~Isberg, U.~Lindstrom, B.~Sundborg and G.~Theodoridis, \emph{{Classical and
  quantized tensionless strings}},
  \href{https://doi.org/10.1016/0550-3213(94)90056-6}{\emph{Nucl. Phys.}
  {\bfseries B411} (1994) 122}
  [\href{https://arxiv.org/abs/hep-th/9307108}{{\ttfamily hep-th/9307108}}].

\bibitem{Bagchi:2013bga}
A.~Bagchi, \emph{{Tensionless Strings and Galilean Conformal Algebra}},
  \href{https://doi.org/10.1007/JHEP05(2013)141}{\emph{JHEP} {\bfseries 05}
  (2013) 141} [\href{https://arxiv.org/abs/1303.0291}{{\ttfamily 1303.0291}}].

\bibitem{Bagchi:2015nca}
A.~Bagchi, S.~Chakrabortty and P.~Parekh, \emph{{Tensionless Strings from
  Worldsheet Symmetries}},
  \href{https://doi.org/10.1007/JHEP01(2016)158}{\emph{JHEP} {\bfseries 01}
  (2016) 158} [\href{https://arxiv.org/abs/1507.04361}{{\ttfamily
  1507.04361}}].

\bibitem{Bagchi:2019cay}
A.~Bagchi, A.~Banerjee and P.~Parekh, \emph{{Tensionless Path from Closed to
  Open Strings}},
  \href{https://doi.org/10.1103/PhysRevLett.123.111601}{\emph{Phys. Rev. Lett.}
  {\bfseries 123} (2019) 111601}
  [\href{https://arxiv.org/abs/1905.11732}{{\ttfamily 1905.11732}}].

\bibitem{Bagchi:2020fpr}
A.~Bagchi, A.~Banerjee, S.~Chakrabortty, S.~Dutta and P.~Parekh, \emph{{A tale
  of three \textemdash{} tensionless strings and vacuum structure}},
  \href{https://doi.org/10.1007/JHEP04(2020)061}{\emph{JHEP} {\bfseries 04}
  (2020) 061} [\href{https://arxiv.org/abs/2001.00354}{{\ttfamily
  2001.00354}}].

\bibitem{Bagchi:2021rfw}
A.~Bagchi, M.~Mandlik and P.~Sharma, \emph{{Tensionless tales: vacua and
  critical dimensions}},
  \href{https://doi.org/10.1007/JHEP08(2021)054}{\emph{JHEP} {\bfseries 08}
  (2021) 054} [\href{https://arxiv.org/abs/2105.09682}{{\ttfamily
  2105.09682}}].

\bibitem{Bagchi:2020ats}
A.~Bagchi, A.~Banerjee and S.~Chakrabortty, \emph{{Rindler Physics on the
  String Worldsheet}},
  \href{https://doi.org/10.1103/PhysRevLett.126.031601}{\emph{Phys. Rev. Lett.}
  {\bfseries 126} (2021) 031601}
  [\href{https://arxiv.org/abs/2009.01408}{{\ttfamily 2009.01408}}].

\bibitem{Bagchi:2021ban}
A.~Bagchi, A.~Banerjee, S.~Chakrabortty and R.~Chatterjee, \emph{{A Rindler
  Road to Carrollian Worldsheets}},
  \href{https://arxiv.org/abs/2111.01172}{{\ttfamily 2111.01172}}.

\bibitem{Bagchi:2019xfx}
A.~Bagchi, A.~Mehra and P.~Nandi, \emph{{Field Theories with Conformal
  Carrollian Symmetry}},
  \href{https://doi.org/10.1007/JHEP05(2019)108}{\emph{JHEP} {\bfseries 05}
  (2019) 108} [\href{https://arxiv.org/abs/1901.10147}{{\ttfamily
  1901.10147}}].

\bibitem{Bagchi:2019clu}
A.~Bagchi, R.~Basu, A.~Mehra and P.~Nandi, \emph{{Field Theories on Null
  Manifolds}}, \href{https://doi.org/10.1007/JHEP02(2020)141}{\emph{JHEP}
  {\bfseries 02} (2020) 141}
  [\href{https://arxiv.org/abs/1912.09388}{{\ttfamily 1912.09388}}].

\bibitem{deBoer:2021jej}
J.~de~Boer, J.~Hartong, N.~A. Obers, W.~Sybesma and S.~Vandoren, \emph{{Carroll
  symmetry, dark energy and inflation}},
  \href{https://arxiv.org/abs/2110.02319}{{\ttfamily 2110.02319}}.

\bibitem{Basu:2018dub}
R.~Basu and U.~N. Chowdhury, \emph{{Dynamical structure of Carrollian
  Electrodynamics}}, \href{https://doi.org/10.1007/JHEP04(2018)111}{\emph{JHEP}
  {\bfseries 04} (2018) 111}
  [\href{https://arxiv.org/abs/1802.09366}{{\ttfamily 1802.09366}}].

\bibitem{Henneaux:2021yzg}
M.~Henneaux and P.~Salgado-Rebolledo, \emph{{Carroll contractions of
  Lorentz-invariant theories}},
  \href{https://doi.org/10.1007/JHEP11(2021)180}{\emph{JHEP} {\bfseries 11}
  (2021) 180} [\href{https://arxiv.org/abs/2109.06708}{{\ttfamily
  2109.06708}}].

\bibitem{Hao:2021urq}
P.-x. Hao, W.~Song, X.~Xie and Y.~Zhong, \emph{{A BMS-invariant free scalar
  model}},  \href{https://arxiv.org/abs/2111.04701}{{\ttfamily 2111.04701}}.

\bibitem{Chen:2021xkw}
B.~Chen, R.~Liu and Y.-f. Zheng, \emph{{On Higher-dimensional Carrollian and
  Galilean Conformal Field Theories}},
  \href{https://arxiv.org/abs/2112.10514}{{\ttfamily 2112.10514}}.

\bibitem{Bidussi:2021nmp}
L.~Bidussi, J.~Hartong, E.~Have, J.~Musaeus and S.~Prohazka, \emph{{Fractons,
  dipole symmetries and curved spacetime}},
  \href{https://arxiv.org/abs/2111.03668}{{\ttfamily 2111.03668}}.

\bibitem{Ciambelli:2018ojf}
L.~Ciambelli and C.~Marteau, \emph{{Carrollian conservation laws and Ricci-flat
  gravity}}, \href{https://doi.org/10.1088/1361-6382/ab0d37}{\emph{Class.
  Quant. Grav.} {\bfseries 36} (2019) 085004}
  [\href{https://arxiv.org/abs/1810.11037}{{\ttfamily 1810.11037}}].

\bibitem{Gupta:2020dtl}
N.~Gupta and N.~V. Suryanarayana, \emph{{Constructing Carrollian CFTs}},
  \href{https://doi.org/10.1007/JHEP03(2021)194}{\emph{JHEP} {\bfseries 03}
  (2021) 194} [\href{https://arxiv.org/abs/2001.03056}{{\ttfamily
  2001.03056}}].

\bibitem{Bergshoeff_2017}
E.~Bergshoeff, J.~Gomis, B.~Rollier, J.~Rosseel and T.~ter Veldhuis,
  \emph{Carroll versus galilei gravity},
  \href{https://doi.org/10.1007/jhep03(2017)165}{\emph{Journal of High Energy
  Physics} {\bfseries 2017} (2017) }.

\bibitem{Ciambelli:2019lap}
L.~Ciambelli, R.~G. Leigh, C.~Marteau and P.~M. Petropoulos, \emph{{Carroll
  Structures, Null Geometry and Conformal Isometries}},
  \href{https://doi.org/10.1103/PhysRevD.100.046010}{\emph{Phys. Rev. D}
  {\bfseries 100} (2019) 046010}
  [\href{https://arxiv.org/abs/1905.02221}{{\ttfamily 1905.02221}}].

\bibitem{Floreanini:1987as}
R.~Floreanini and R.~Jackiw, \emph{{Selfdual Fields as Charge Density
  Solitons}}, \href{https://doi.org/10.1103/PhysRevLett.59.1873}{\emph{Phys.
  Rev. Lett.} {\bfseries 59} (1987) 1873}.

\bibitem{Sonnenschein:1988ug}
J.~Sonnenschein, \emph{{CHIRAL BOSONS}},
  \href{https://doi.org/10.1016/0550-3213(88)90339-2}{\emph{Nucl. Phys. B}
  {\bfseries 309} (1988) 752}.

\bibitem{Casali:2016atr}
E.~Casali and P.~Tourkine, \emph{{On the null origin of the ambitwistor
  string}}, \href{https://doi.org/10.1007/JHEP11(2016)036}{\emph{JHEP}
  {\bfseries 11} (2016) 036}
  [\href{https://arxiv.org/abs/1606.05636}{{\ttfamily 1606.05636}}].

\bibitem{Casali:2017zkz}
E.~Casali, Y.~Herfray and P.~Tourkine, \emph{{The complex null string, Galilean
  conformal algebra and scattering equations}},
  \href{https://doi.org/10.1007/JHEP10(2017)164}{\emph{JHEP} {\bfseries 10}
  (2017) 164} [\href{https://arxiv.org/abs/1707.09900}{{\ttfamily
  1707.09900}}].

\bibitem{Grumiller:2019tyl}
D.~Grumiller and W.~Merbis, \emph{{Near horizon dynamics of three dimensional
  black holes}},
  \href{https://doi.org/10.21468/SciPostPhys.8.1.010}{\emph{SciPost Phys.}
  {\bfseries 8} (2020) 010} [\href{https://arxiv.org/abs/1906.10694}{{\ttfamily
  1906.10694}}].

\bibitem{Deser:1982vy}
S.~Deser, R.~Jackiw and S.~Templeton, \emph{{Three-Dimensional Massive Gauge
  Theories}}, \href{https://doi.org/10.1103/PhysRevLett.48.975}{\emph{Phys.
  Rev. Lett.} {\bfseries 48} (1982) 975}.

\bibitem{Li:2008dq}
W.~Li, W.~Song and A.~Strominger, \emph{{Chiral Gravity in Three Dimensions}},
  \href{https://doi.org/10.1088/1126-6708/2008/04/082}{\emph{JHEP} {\bfseries
  04} (2008) 082} [\href{https://arxiv.org/abs/0801.4566}{{\ttfamily
  0801.4566}}].

\bibitem{Grumiller:2008qz}
D.~Grumiller and N.~Johansson, \emph{{Instability in cosmological topologically
  massive gravity at the chiral point}},
  \href{https://doi.org/10.1088/1126-6708/2008/07/134}{\emph{JHEP} {\bfseries
  07} (2008) 134} [\href{https://arxiv.org/abs/0805.2610}{{\ttfamily
  0805.2610}}].

\bibitem{Maloney:2009ck}
A.~Maloney, W.~Song and A.~Strominger, \emph{{Chiral Gravity, Log Gravity and
  Extremal CFT}}, \href{https://doi.org/10.1103/PhysRevD.81.064007}{\emph{Phys.
  Rev. D} {\bfseries 81} (2010) 064007}
  [\href{https://arxiv.org/abs/0903.4573}{{\ttfamily 0903.4573}}].

\bibitem{Bagchi:2018ryy}
A.~Bagchi, R.~Basu, S.~Detournay and P.~Parekh, \emph{{Flatspace Chiral
  Supergravity}}, \href{https://doi.org/10.1103/PhysRevD.97.106020}{\emph{Phys.
  Rev.} {\bfseries D97} (2018) 106020}
  [\href{https://arxiv.org/abs/1801.03245}{{\ttfamily 1801.03245}}].

\bibitem{Bagchi:2016yyf}
A.~Bagchi, S.~Chakrabortty and P.~Parekh, \emph{{Tensionless Superstrings: View
  from the Worldsheet}},
  \href{https://doi.org/10.1007/JHEP10(2016)113}{\emph{JHEP} {\bfseries 10}
  (2016) 113} [\href{https://arxiv.org/abs/1606.09628}{{\ttfamily
  1606.09628}}].

\bibitem{Bagchi:2017cte}
A.~Bagchi, A.~Banerjee, S.~Chakrabortty and P.~Parekh, \emph{{Inhomogeneous
  Tensionless Superstrings}},
  \href{https://doi.org/10.1007/JHEP02(2018)065}{\emph{JHEP} {\bfseries 02}
  (2018) 065} [\href{https://arxiv.org/abs/1710.03482}{{\ttfamily
  1710.03482}}].

\bibitem{Bagchi:2018wsn}
A.~Bagchi, A.~Banerjee, S.~Chakrabortty and P.~Parekh, \emph{{Exotic Origins of
  Tensionless Superstrings}},
  \href{https://doi.org/10.1016/j.physletb.2019.135139}{\emph{Phys. Lett. B}
  (2019) } [\href{https://arxiv.org/abs/1811.10877}{{\ttfamily 1811.10877}}].

\bibitem{Lodato:2016alv}
I.~Lodato and W.~Merbis, \emph{{Super-BMS$_{3}$ algebras from $ \mathcal{N}=2 $
  flat supergravities}},
  \href{https://doi.org/10.1007/JHEP11(2016)150}{\emph{JHEP} {\bfseries 11}
  (2016) 150} [\href{https://arxiv.org/abs/1610.07506}{{\ttfamily
  1610.07506}}].

\bibitem{Bagchi:2022owq}
A.~Bagchi, D.~Grumiller and P.~Nandi, \emph{{Carrollian superconformal theories
  and super BMS}},  \href{https://arxiv.org/abs/2202.01172}{{\ttfamily
  2202.01172}}.

\bibitem{Bonelli:2003kh}
G.~Bonelli, \emph{{On the tensionless limit of bosonic strings, infinite
  symmetries and higher spins}},
  \href{https://doi.org/10.1016/j.nuclphysb.2003.07.002}{\emph{Nucl. Phys.}
  {\bfseries B669} (2003) 159}
  [\href{https://arxiv.org/abs/hep-th/0305155}{{\ttfamily hep-th/0305155}}].

\bibitem{Lindstrom:2003mg}
U.~Lindstrom and M.~Zabzine, \emph{{Tensionless strings, WZW models at critical
  level and massless higher spin fields}},
  \href{https://doi.org/10.1016/j.physletb.2004.01.035}{\emph{Phys. Lett.}
  {\bfseries B584} (2004) 178}
  [\href{https://arxiv.org/abs/hep-th/0305098}{{\ttfamily hep-th/0305098}}].

\end{thebibliography}\endgroup
\end{document}